\documentclass[a4paper,11pt]{article}
\pdfoutput=1 

\usepackage{jheppub} 

\usepackage[T1]{fontenc} 
\usepackage{enumitem}
\usepackage{tikz}
\usetikzlibrary{decorations.pathmorphing}
\usepackage{subcaption} 
\usepackage{caption}
\usepackage{tkz-tab}
\usepackage{amsmath}
\usepackage{amssymb}
\usepackage{amsfonts}
\usepackage{amsthm}
\usepackage{mathrsfs}
\usepackage{comment}
\usepackage{graphicx}
\usepackage{epstopdf}

\def\be {\begin{equation}}
\def\ee {\end{equation}}
\def\bea {\begin{eqnarray}}
\def\eea {\end{eqnarray}}
\def\nn {\nonumber}

\title{Dynamical black holes in low-energy string theory}

\author[a]{Pedro Aniceto}
\author[b,1]{Jorge V. Rocha,\note{Corresponding author.}}

\affiliation[a]{Departamento de Matem\'atica, Instituto Superior T\'ecnico, 
Universidade de Lisboa,\\Avenida~Rovisco Pais 1, 1049 Lisboa, Portugal}
\affiliation[b]{Departament de F\'isica Qu\`antica i Astrof\'isica, Institut de Ci\`encies del Cosmos (ICCUB),\\
Universitat de Barcelona, Mart\'i i Franqu\`es 1, E-08028 Barcelona, Spain}

\emailAdd{pedro.aniceto@tecnico.ulisboa.pt}
\emailAdd{jvrocha@icc.ub.edu}

\abstract{We investigate time-dependent spherically symmetric solutions of the four-dimensional Einstein-Maxwell-axion-dilaton system, with the dilaton coupling that occurs in low-energy effective heterotic string theory. A class of dilaton-electrovacuum radiating solutions with a trivial axion, previously found by G\"uven and Y\"or\"uk, is re-derived in a simpler manner and its causal structure is clarified. It is shown that such dynamical spacetimes featuring apparent horizons do not possess a regular light-like past null infinity or future null infinity, depending on whether they are radiating or accreting.
These solutions are then extended in two ways.
First we consider a Vaidya-like generalisation, which introduces a null dust source.
Such spacetimes are used to test the status of cosmic censorship in the context of low-energy string theory. We prove that---within this family of solutions---regular black holes cannot evolve into naked singularities by accreting null dust, unless standard energy conditions are violated.
Secondly, we employ S-duality to derive new time-dependent dyon solutions with a nontrivial axion turned on. Although they share the same causal structure as their Einstein-Maxwell-dilaton counterparts, these solutions possess both electric and magnetic charges.}

\begin{document} 
\maketitle
\flushbottom

\section{Introduction}\label{sec:intro}

Low-energy effective string theories arising from various compactification schemes and from different progenitor string theories come equipped with multiple scalar fields~\cite{Green:1987mn,Ortin:2004ms}. Among these, the dilaton, whose expectation value determines the string coupling, is a prime example. But many other (pseudo)scalar fields can be easily obtained from form fields when the compactification down to four dimensions involves a manifold with closed cycles, as is typically the case. Vector fields also naturally occur in the bosonic sector of such 4D effective theories, in addition to gravity and scalar fields, and the coupling with the latter takes a characteristic exponential form.

All these main features of low-energy string theory are displayed in a simple model theory, the Einstein-Maxwell-axion-dilaton (EMad) system, which we adopt in the present paper. It describes two massless scalar fields with distinct couplings to the abelian vector field: the dilaton couples to the square of the Maxwell field strength $F^2$, whereas the axion couples to the topological term $F^*F$ and is therefore a pseudoscalar. Apart from its simplicity, this model is useful also as a consistent truncation of the low-energy effective heterotic superstring compactified to four dimensions~\cite{Duff:1996qp}.

This article concerns black hole solutions in such four-dimensional effective theories. It is well known that the presence of scalar fields affects some of their basic properties~\cite{Boulware:1986dr}. By investigating static, spherically symmetric solutions of the Einstein-Maxwell-dilaton (EMd) system, Gibbons and Maeda~\cite{Gibbons:1987ps} and Garfinkle, Horowitz and Strominger~\cite{Garfinkle:1990qj} found that the inclusion of the dilaton field shifts the critical value of the electromagnetic charge that separates black hole geometries from naked singularities, when compared with the Reissner-Nordstr\"om solution. It also causes the would-be inner horizon to be a singular surface, which in addition becomes spacelike.

These static configurations have become known as GMGHS solutions, and they fall in the broader class of dilaton black holes. They possess either a purely magnetic or purely electric Maxwell field and have a vanishing axion. On the other hand, dyon solutions---carrying simultaneously electric and magnetic charges---require a nontrivial axion, since this field is sourced by the topological term $\epsilon_{\mu\nu\rho\sigma}F^{\mu\nu} F^{\rho\sigma}$.
Indeed, static dyon solutions were presented shortly after in Ref.~\cite{Shapere:1991ta}, but only for an EMad theory with a particular dilaton coupling. When the coupling parameter takes this special value, an $SL(2,\mathbb{R})$ symmetry group---known as S-duality---emerges, which was then exploited to construct dyonic solutions.
Dyon solutions of the simpler EMd system (without the topological term for the Maxwell field) are also known~\cite{Kallosh:1992ii,Horowitz:1992jp,Goulart:2016cuv} and they differ from the previous ones even at the level of the causal structure.
Finally, a multitude of charged extended black holes ($p$-brane solutions) in 10D low-energy string theory can be obtained from the static solutions of~\cite{Gibbons:1987ps,Garfinkle:1990qj} by a simple uplifting procedure, as has been discussed in Ref.~\cite{Horowitz:1991cd}.

Going back to the GMGHS solutions, they are known to be classically stable against gravitational, electromagnetic and dilaton field perturbations~\cite{Holzhey:1991bx}.
Their extension to include rotation was studied perturbatively around the static solutions in Ref.~\cite{Horne:1992zy} but exact solutions in four-dimensional EMd theory with finite angular momentum are only known for the dilaton coupling corresponding to dimensional reduction of pure gravity on a circle from 5D to 4D~\cite{Rasheed:1995zv}\footnote{Rotating charged black holes have been obtained also for other values of the dilaton coupling, namely for toroidally compactified heterotic string theory~\cite{Sen:1992ua,Cvetic:1996kv}, but they include additional matter fields.}.
In five dimensions, charged rotating black holes were also obtained for particular values of the dilaton coupling~\cite{Cvetic:1996xz,Matos:1996km} and rotating black ring solutions carrying dipole magnetic charge have been constructed~\cite{Emparan:2004wy,Rocha:2011vv}.
Uniqueness results for static solutions in EMd theory were proven in Refs.~\cite{Mars:2001pz,Gibbons:2002ju}, and later this was extended to the stationary case in Ref.~\cite{Yazadjiev:2010bj}.

All the configurations discussed above are stationary. The GMGHS solutions in particular are static and spherically symmetric. In Einstein(-Maxwell) theory the assumption of spherical symmetry implies the staticity of (electro-)vacuum solutions, as a consequence of Birkhoff's theorem (and Gauss' law). However, in the presence of scalar fields, time-dependence is not excluded since there can exist scalar radiation even under the assumption of spherical symmetry.

In fact, radiative spherically symmetric solutions in EMd theory do exist. They were found by G\"uven and Y\"or\"uk~\cite{Gueven:1996zm} as a special subclass of Robinson-Trautman solutions. These solutions have constant electric charge, approach static solutions either in the far past or in the far future and, curiously, are known only for the special dilaton coupling that exhibits the emergent $SL(2,\mathbb{R})$ duality. Their existence is made possible by evading Birkhoff theorem's assumptions~\cite{Faraoni:2010rt}, namely the Maxwell and dilaton energy-momentum tensors are time-dependent. They can be regarded as a generalization of the well-known Roberts' self-similar solution for the Einstein-massless scalar system~\cite{Roberts:1989sk,Roberts:1999ui} to include a Maxwell field. More recently, this solution was embedded in ${\cal N}=4$ gauged supergravity~\cite{Lu:2014eta} and similar time-dependent solutions with supergravity-inspired potentials for the scalar field have been discussed in~\cite{Zhang:2014sta,Zhang:2014dfa,Fan:2016yqv}.

A particularity of the G\"uven-Y\"or\"uk solutions is that, although the mass and dilaton charge evolve in time, their product is determined by the electric charge, which is kept constant. A natural question to ask then is whether these radiating solutions lead to violations of the weak cosmic censorship conjecture~\cite{Penrose:1969pc,Wald:1997wa}. I.e., can a regular, asymptotically flat, black hole in low-energy effective string theory evolve into a naked singularity by radiating sufficient energy or by accreting physically reasonable matter?

In Einstein-Maxwell theory this question was answered in the negative by Sullivan and Israel~\cite{Sullivan:1980}, who demonstrated the impossibility of overcharging a spherical black hole by bombarding it with charged null dust satisfying the standard energy conditions. In that case the evolution of the system was described by the Bonnor-Vaidya solution~\cite{Bonnor:1970zz}, but the result is in line with an earlier study based on an attempt to overcharge Reissner-Nordstr\"om black holes with thin shells~\cite{Boulware:1973}. Similarly, weak cosmic censorship has prevailed in recent attempts to overspin black holes (in higher dimensions) with rotating thin shells~\cite{Delsate:2014iia,Rocha:2015tda}.

At first sight, radiating spacetimes in Einstein-Maxwell-dilaton would appear to favour weak cosmic censorship violation, since the mass is continuously decreasing and at some point extremality will be exceeded. Potential violations of weak cosmic censorship in Einstein-Maxwell-dilaton theory were suggested over two decades ago in Ref.~\cite{Horne:1993sy} by studying extreme dilatonic black holes with a positive cosmological constant. In the present paper we consider strictly zero cosmological constant.

In a previous paper~\cite{Aniceto:2015klq} a possible violation of cosmic censorship was considered in EMd theory by bombarding black holes with a radial stream of charged null fluids. Indeed, it was found that the upper bound on the mass of a regular static black hole can be reached by accreting such matter, while still satisfying usual energy conditions. However, this is not enough to guarantee the resulting spacetime features a naked singularity. The reason is that the extremality condition being violated applies only to the static solutions, whereas the accreting spacetime is time-dependent. 

To fully address cosmic censorship in EMd theory one therefore needs to understand the causal structure of their time-dependent solutions, namely the G\"uven-Y\"or\"uk (GY) black holes. A dedicated analysis of the causal structure of these spacetimes appears to be missing in the literature, so our first goal is to perform such a detailed study. In this respect, we find that
\begin{itemize}
\item  The GY geometries are actually not asymptotically flat, in the sense that they do not display the entire Minkowski asymptotic structure: depending on whether one considers radiating or accreting solutions the spacetime does not possess either a light-like past null infinity ${\cal J}^-$ or future null infinity ${\cal J}^+$. The origin of this peculiarity is tied to the fact that the spacetime radiates or accretes forever. Instead, those solutions that have an apparent horizon seem to describe a Big-Bang-like singularity evolving into a black hole or, by taking its time reverse, a white hole evolving into a Big-Crunch-like singularity;
\item  Within the class of dilaton-electrovacuum spacetimes captured by the GY solution, the character of the curvature singularity does not change throughout time evolution.
\end{itemize}

The latter point shows that unmodified GY solutions cannot evolve a black hole into a naked singularity. We then consider the addition of a (neutral) null dust source, equivalent to the Vaidya spacetime~\cite{Vaidya:1951zza,Vaidya:1951zz,Lindquist:1965zz} in general relativity or to the Bonnor-Vaidya solution~\cite{Bonnor:1970zz} in Einstein-Maxwell. This can be easily accomplished by relaxing a constraint, which must be imposed to obtain vacuum solutions. Then, a simple proof is given to the effect that
\begin{itemize}
\item The resulting time-dependent spacetimes---sourced by null dust---comply with the weak cosmic censorship conjecture. Evolution of a black hole into a naked singularity requires the additional matter source to violate standard energy conditions.
\end{itemize}

Finally, it is striking that the S-duality in the Einstein-Maxwell-axion-dilaton theory is available {\em precisely} for the value of the coupling constant for which time-dependent solutions are explicitly known. Thus, by taking the GY solutions as seeds and performing $SL(2,\mathbb{R})$ transformations
\begin{itemize}
\item  We derive novel radiating dyonic solutions with a nontrivial axion field. Their line element is exactly the same as for the G\"uven-Y\"or\"uk solution so they share the same causal structure, but they possess both electric and magnetic charges.
\end{itemize}

\bigskip
The rest of the paper is organised as follows.
In Section~\ref{sec:fieldeqs} the Lagrangian describing the Einstein-Maxwell-dilaton system is presented, together with the corresponding field equations. Section~\ref{sec:static} briefly reviews the static GMGHS solutions. Section~\ref{sec:Tdependent} is dedicated to a detailed analysis of the radiating vacuum solutions originally found by G\"uven and Y\"or\"uk. The inclusion of a null dust component is considered in Section~\ref{sec:NullDust}, where it is also proved that none of such solutions can violate the weak cosmic censorship conjecture. This conclusion is also reached in Section~\ref{sec:matching} by examining the limiting case when the null dust is compressed to an infinitely thin shell. This is studied by matching two G\"uven-Y\"or\"uk spacetimes across null hypersurfaces. In Section~\ref{sec:axion} we consider the more general Einstein-Maxwell-axion-dilaton theory and derive dyonic generalisations of the GY solutions, which are also spherically symmetric and time-dependent. 

The paper is complemented with three appendices. Appendix~\ref{App1} presents a simple derivation of the GY solutions, indicating why a straightforward generalisation to other values of the dilaton coupling is not valid. Appendix~\ref{App2} discusses a special class of (pathological) GY solutions for which the mass and dilaton charge evolve in (retarded/advanced) time according to power laws. Details of the formalism for matching spacetimes using null surface layers are given in Appendix~\ref{App3}.

\section{The Einstein-Maxwell-dilaton system}\label{sec:fieldeqs}

We consider the following Lagrangian for Einstein-Maxwell-dilaton theory (henceforth we adopt geometrised units for which the gravitational constant and the speed of light are $G=c=1$),
\be
{\cal L} =\frac{\sqrt{-g}}{16\pi}\left[R-2\nabla_\mu\phi \nabla^\mu\phi-e^{-2\phi}F_{\mu\nu}F^{\mu\nu}\right]\,,\label{lagrangian}
\ee
where $R$ denotes the Ricci scalar, $\phi$ the dilaton scalar field, and $F_{\mu\nu}$ stands for the electromagnetic field strength.
Here, $g$ represents the determinant of the metric $g_{\mu\nu}$, which is used to raise and lower tensor indices.
The field equations derived from Eq.~\eqref{lagrangian} read 
\begin{subequations}\label{eq:fieldeqs}
\bea
&&\nabla_\mu \left( e^{-2 \phi} F^{\mu \nu} \right) = 0\,,\label{MaxwellEOM}\\
&&\nabla^2 \phi + \frac{1}{2}e^{-2\phi} F_{\mu \nu} F^{\mu \nu} = 0\,,\label{dilatonEOM}\\
&&G_{\mu\nu} = 8\pi  T_{\mu\nu} \equiv 8\pi \left(T_{\mu\nu}^{\rm (dil)} + e^{-2\phi}T_{\mu\nu}^{\rm (EM)} \right)\,,\label{EinsteinEOM}
\eea
\end{subequations}
where $G_{\mu\nu} = R_{\mu\nu} - \frac{1}{2}R g_{\mu\nu}$ is the Einstein tensor.
We have split the total energy-momentum tensor $T_{\mu\nu}$ into two pieces according to their different origins: a contribution from the dilaton,
\be
8\pi T_{\mu\nu}^{\rm (dil)} = 2\nabla_\mu\phi\nabla_\nu\phi - g_{\mu\nu}(\nabla\phi)^2\,, \label{Tdil}
\ee
and a contribution from the electromagnetic field,
\be
8\pi T_{\mu\nu}^{\rm (EM)} = \left( 2F_{\mu \alpha} {F_\nu}^\alpha -\frac{1}{2}g_{\mu \nu}F^2 \right)\,.
\label{TEM}
\ee
Later, in Section~\ref{sec:NullDust}, we will include an additional contribution in the energy-momentum tensor~\eqref{EinsteinEOM} to account for the possibility of a null dust source. In Section~\ref{sec:axion} we will add the axion field to obtain the full Einstein-Maxwell-axion-dilaton system.

\section{Static black hole solutions}\label{sec:static}

Now we briefly review static solutions of the field equations~\eqref{eq:fieldeqs}, which were found in Refs.~\cite{Gibbons:1987ps,Garfinkle:1990qj}.
The electrically charged solutions read
\begin{eqnarray}
 ds^2 &=& -\left(1-\frac{2M}{r}\right) dt^2 + \left(1-\frac{2M}{r}\right)^{-1} dr^2 + r^2 \left(1-\frac{2D}{r}\right) d\Omega^2 \,, \label{metricGHS} \\
 F &=&-\frac{Q}{r^2} dt\wedge dr\,,\label{FaradayGHS} \\
 e^{2\phi(r)}&=& e^{2\phi_0}\left(1-\frac{2D}{r}\right)\,, \label{dilatonTdependent} 
\end{eqnarray}
where $d\Omega^2 = d\theta^2 + \sin^2\theta d\varphi^2$ denotes the line element of the unit 2-sphere.
The parameters $M, Q$ and $D$ represent the physical mass, the electric charge, and the dilatonic charge, respectively. However, note they are not all independent, as they are constrained by~\cite{Garfinkle:1990qj}
\be
e^{-2\phi_0} Q^2 = 2MD\,.
\label{eq:QMDconstraint}
\ee
Therefore, these black holes are said to feature scalar hair of the secondary kind. The scalar charge is computed by an integral over the 2-sphere at spatial infinity,
\be
D=\frac{1}{4\pi} \int_{S^2_\infty} d^2\Sigma^\mu \nabla_\mu \phi\,.
\ee
The free parameter $\phi_0$ is a real constant that determines the asymptotic value of the dilaton field.

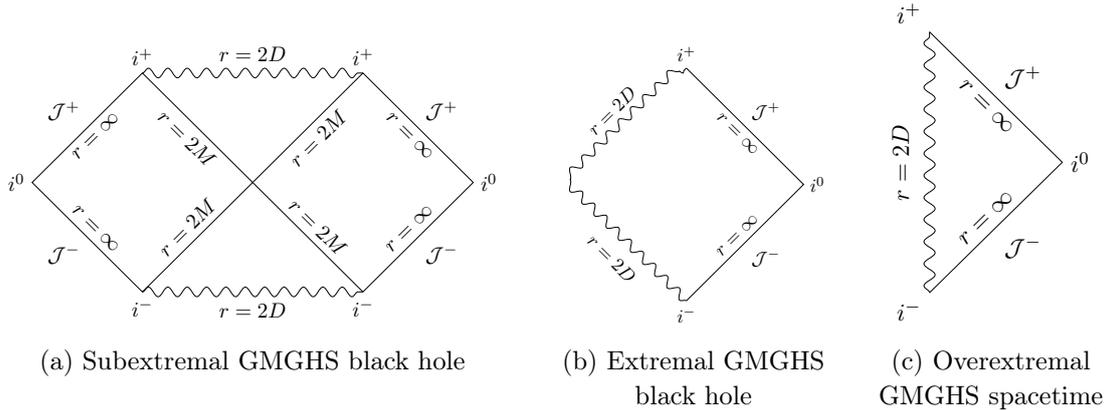
\begin{figure}[!t]
	\centering
	\begin{subfigure}[t]{0.45\textwidth}
		\centering
		\resizebox{\linewidth}{!}{
			\begin{tikzpicture}
			\node (I)    at ( 2,0)  {} ;
			\node (II)   at (-2,0)  {};
			\node (III)  at (0, 1) {};
			\node (IV)   at (0,-1) {};

			\path  
			(II) +(90:2)  coordinate[label=90:$i^+$]  (IItop)
			+(-90:2) coordinate[label=-90:$i^-$] (IIbot)
			+(0:2)   coordinate                  (IIright)
			+(180:2) coordinate[label=180:$i^0$] (IIleft)
			;
			\draw (IIleft) -- 
			node[midway, above left]    {$\cal{J}^+$}
			node[midway, below, sloped] {$r=\infty$}
			(IItop) --
			node[midway, below, sloped] {$r=2M$}
			(IIright) -- 
			node[midway, above, sloped] {$r=2M$}
			(IIbot) --
			node[midway, below left]    {$\cal{J}^-$}   
			node[midway, above, sloped] {$r=\infty$} 
			(IIleft) -- cycle;
			
			\path 
			(I) +(90:2)  coordinate[label=90:$i^+$]  (Itop)
			+(-90:2) coordinate[label=-90:$i^-$]  (Ibot)
			+(180:2) coordinate (Ileft)
			+(0:2)   coordinate[label=0:$i^0$]  (Iright)
			;
			\draw  (Ileft) -- 
			node[midway, below, sloped] {$r=2M$}
			(Itop) -- 
			node[midway, above right]    {$\cal{J}^+$}
			node[midway, below, sloped] {$r=\infty$}
			(Iright) -- 
			node[midway, below right]    {$\cal{J}^-$}
			node[midway, above, sloped] {$r=\infty$}
			(Ibot) -- 
			node[midway, above, sloped] {$r=2M$}
			(Ileft) -- 
			cycle;
			
			\draw[decorate,decoration=snake] (IItop) -- (Itop)
			node[midway, above, inner sep=2mm] {$r=2D$};
			
			\draw[decorate,decoration=snake] (IIbot) -- (Ibot)
			node[midway, below, inner sep=2mm] {$r=2D$};
			
			\end{tikzpicture}
		}
		\caption{Subextremal GMGHS black hole}
		\label{fig:subfig1}
	\end{subfigure}
	\quad
		\begin{subfigure}[t]{0.25 \textwidth}
			\centering
			\resizebox{\linewidth}{!}{
				\begin{tikzpicture}
				\node (I)    at (1,0)   {};
				\node (II)    at (1,0)   {};
				
				\path
				(II) +(90:2.4)  coordinate[label=90:$i^+$]  (IItop)
				+(-90:2.4) coordinate[label=-90:$i^-$] (IIbot)
				+(0:2.4) coordinate[label=0:$i^0$] (IIright)
				+(180:2.4) coordinate[label=180:$ $] (IIleft)
				;

				\draw[decorate,decoration=snake] (IIbot) -- 
				node[midway, below, sloped, outer sep=2mm] {$r=2D$}
				(IIleft) --
				node[midway, above, sloped, outer sep=2mm] {$r=2D$}
				(IItop);
				
				\draw (IItop) -- 
				node[midway, above right]    {$\cal{J}^+$}
				node[midway, below, sloped] {$r=\infty$}
				(IIright) --
				node[midway, below right] {$\cal{J}^-$}
				node[midway, above, sloped] {$r=\infty$}
				(IIbot) ;
				\end{tikzpicture}
			}
			\caption{\centering Extremal GMGHS black hole}   
			\label{fig:subfig_extremal}
		\end{subfigure} 
	\quad
	\begin{subfigure}[t]{0.20 \textwidth}
		\centering
		\resizebox{\linewidth}{!}{
			\begin{tikzpicture}
			\node (I)    at (1,0)   {};

			\path  
			(I) +(120:2.4)  coordinate[label=120:$i^+$]  (Itop)
			+(-120:2.4) coordinate[label=-120:$i^-$] (Ibot)
			+(0:0.9) coordinate[label=0:$i^0$] (Ileft)
			;
			
			\draw[decorate,decoration=snake] (Ibot) -- (Itop)
			node[midway, above, sloped, outer sep=2mm] {$r=2D$};
			
			\draw (Itop) -- 
			node[midway, above right]    {$\cal{J}^+$}
			node[midway, below, sloped] {$r=\infty$}
			(Ileft) --
			node[midway, below right] {$\cal{J}^-$}
			node[midway, above, sloped] {$r=\infty$}
			(Ibot) ;
			\end{tikzpicture}
		}
		\caption{\centering Overextremal GMGHS spacetime}   
		\label{fig:subfig2}
	\end{subfigure} 
\caption[Penrose diagrams of GMGHS]{Carter-Penrose diagrams representing the causal structure of GMGHS solutions for (a) the subextremal case $M>D$, (b) the extremal case $M=D$, and (c) the superextremal case $M<D$.} 
	\label{fig:Penrose1}
\end{figure}

The spherical surface $r=2D$ is singular. It also has vanishing area, signalling that spacetime ends there.
If $M>D$ there will be an event horizon at $r=2M$. When $M<D$ the geometry corresponds to a naked singularity and for $M=D$ the singular surface is light-like. The causal structure of these solutions is made explicit in the Carter-Penrose diagrams displayed in figure~\ref{fig:Penrose1}.

It will be convenient to express the solution above in retarded/advanced Eddington-Finkelstein coordinates. This is accomplished by defining $u\equiv t-\epsilon r_*$, where $dr_*/dr=\left(1-2M/r\right)^{-1}$ and $\epsilon=+1$ ($\epsilon=-1$) yields the retarded (advanced) null coordinate. Then the solution becomes
\begin{eqnarray}
 ds^2 &=& -\left(1-\frac{2M}{r}\right) du^2 - \epsilon du dr + r^2 \left(1-\frac{2D}{r}\right) d\Omega^2 \,, \\
 F &=&-\frac{Q}{r^2} du\wedge dr\,, \\
 e^{2\phi(r)}&=& e^{2\phi_0}\left(1-\frac{2D}{r}\right)\,.
\end{eqnarray}
Finally, changing the radial coordinate to the areal radius according to
\be
R^2\equiv r(r-2D)\,,
\label{eq:Rareal}
\ee
we obtain
\begin{eqnarray}
 ds^2 &=& -\left(1-\frac{2M}{D+\sqrt{R^2+D^2}}\right) du^2 - \epsilon \frac{R}{\sqrt{R^2+D^2}} du dR + R^2 d\Omega^2 \,, \label{eq:GHS_areal}\\
 F &=&-\frac{Q R}{\sqrt{R^2+D^2}(D+\sqrt{R^2+D^2})^2} du\wedge dR\,, \\
 e^{2\phi(R)}&=& e^{2\phi_0}\left(1-\frac{2D}{D+\sqrt{R^2+D^2}}\right)\,.
\end{eqnarray}

For simplicity, in what follows we shall always set the asymptotic value of the dilaton to zero, $\phi_0=0$. A nonvanishing $\phi_0$ can always be generated by a trivial shift $\phi\to\phi+\phi_0$.\footnote{This transformation leaves the Lagrangian~\eqref{lagrangian} invariant when simultaneously accompanied by a rescaling of the gauge field.}

\section{Time-dependent vacuum solutions with constant charge}\label{sec:Tdependent}

In this section we begin by presenting non-static---but spherically symmetric---solutions of the Einstein-Maxwell-dilaton field equations without extra matter sources. Although the spacetime is dynamical, the electromagnetic field is time-independent.

These solutions were first obtained in Ref.~\cite{Gueven:1996zm} as a special class of Robinson-Trautman solutions. Nevertheless, a somewhat simpler derivation is possible by following the approach originally introduced by Vaidya~\cite{Vaidya:1951zza}. In short, one takes a static solution as seed but allows the mass and dilatonic charge parameters to depend on the retarded (or advanced) time coordinate $u$. In general, this will introduce nonvanishing sources in the field equations. However, they are all satisfied without sources if the $\{uu\}$ component of the metric is simultaneously shifted by a particular function of the null coordinate $u$.
In the Appendix we present this derivation without assuming a particular value for the strength of the dilaton coupling. I.e., we allow for a more general coupling term in the Lagrangian~\eqref{lagrangian} of the form $e^{2a\phi}F^2$. However, we find that the procedure is successful only if $a^2=1$.\footnote{Note that the case $a=-1$ can be trivially obtained from the theory with $a=1$ by reflection, $\phi\to-\phi$.}

The solution obtained by following this method, for the theory with $a=1$, is
\begin{subequations}\label{GYsolution}
\bea
 ds^2 &=&-\left(1-\frac{2M(u)}{r}-2\epsilon D'(u)\right) du^2 - 2\epsilon\, du\, dr + r^2 \left(1-\frac{2D(u)}{r}\right) d\Omega^2\,, \label{eq:metricsol}\\
 F &=& -\frac{Q}{r^2} du\wedge dr\,,\\
 e^{2\phi}&=&\left(1-\frac{2D(u)}{r}\right)\,,
\eea
\end{subequations}
with the constraints
\bea
&&2M(u)D(u) = Q^2 = {\rm const.}\,, \\
&&2 D(u)^3 D''(u)+\epsilon Q^2 D'(u)=0\,.
\label{eq:ODEforD}
\eea
Recall that $\epsilon=+1$ or $\epsilon=-1$, corresponding to a radiating or an accreting spacetime, respectively.
This is almost identical to what one would get by simply taking the static GMGHS spacetime and promoting $M$ and $D$ to be functions of the retarded/advanced time $u$. The only difference lies in the additional term $-2\epsilon D'(u)$ appearing in the metric component $g_{uu}$.

Note also that $D$ must satisfy the non-linear ODE~\eqref{eq:ODEforD}.
This condition must be imposed to avoid additional matter sources and it is trivially solved by having $M, D$ and $Q$ being constants, which of course just returns the GMGHS static solution.

Another possibility, in the absence of an electromagnetic field ($Q=0$) but with a nontrivial dilaton, is
\be
M=0 \qquad {\rm and} \qquad D'={\rm const.}\,,
\ee
in which case we recover the well known self-similar solution found by Roberts~\cite{Roberts:1989sk} for the Einstein-(massless) scalar system. In other words, Eqs.~\eqref{GYsolution} represent a charged generalization of Roberts' solution to low-energy heterotic string theory.

In addition to the above two cases, we have non-trivial solutions whenever the ODE~\eqref{eq:ODEforD} is satisfied. Such solutions are discussed in detail in subsection~\ref{sec:vacuum}.

It is useful to perform a coordinate transformation to express the solutions using the areal radius, as defined in Eq.~\eqref{eq:Rareal}.
The dynamical solutions~\eqref{GYsolution} then take the general form
\begin{subequations}\label{eq:ansatz}
\bea
 ds^2 &=&-A(u,R) du^2 - 2\epsilon W(u,R)\, du dR + R^2 d\Omega^2\,,  \label{eq:ansatz1}\\
 F &=& -\frac{X(u,R)}{r^2} du\wedge dR\,, \label{eq:ansatz2}\\
 \phi&=&\phi(u,R)\,, \label{eq:ansatz3}
\eea 
\end{subequations}
with the following expressions for the intervening functions:
\bea
A(u,R) &=& 1 - \frac{2 M(u)}{\sqrt{D(u)^2+R^2}+D(u)} + 2\epsilon D(u) D'(u)\frac{W(u,R)}{R}\,, \label{A_function_areal}\\
W(u,R) &=& \frac{R}{\sqrt{D(u)^2+R^2}}\,, \\
X(u,R) &=& \frac{Q\, R^3}{\sqrt{D(u)^2+R^2} \left(\sqrt{D(u)^2+R^2}+D(u)\right)^2}\,, \\
\phi(u,R) &=& \frac{1}{2} \log \left(1-\frac{2 D(u)}{\sqrt{D(u)^2+R^2}+D(u)}\right)\,. \label{dilaton_field_areal}
\eea

The departure of the line element from Minkowski spacetime ($ds^2=\eta_{\mu\nu}dx^\mu dx^\nu$) is controlled by just two nonvanishing differences,
\be
h_{uu}\equiv g_{uu}-\eta_{uu}= 1-A\,, \qquad\qquad
h_{uR}\equiv g_{uR}-\eta_{uR}= \epsilon(1-W)\,.
\ee
For large radii, they both fall off at least as fast as $R^{-1}$. Therefore their radial derivatives also decay as required for asymptotic flatness in the weak sense of \cite{Bondi:1962px,Sachs:1962wk}, namely
\be
\partial_R h_{\mu\nu}=O(R^{-2})\,, \qquad\qquad
\partial_R^2 h_{\mu\nu}=O(R^{-3})\,.
\ee
However, $u$-derivatives do not fall off sufficiently fast to guarantee a finite expression for the energy. For example,
\be
\partial_u h_{uu}  = \frac{-2\epsilon D D''-2\epsilon D'^2+2 M'}{R} + O(R^{-2})
= -\frac{2\epsilon D'^2}{R} + O(R^{-2})\,. \label{eq:du_huu}
\ee
In this sense the solution cannot be regarded as asymptotically flat.
We will confirm this below by studying the behavior of null geodesics and finding that the spacetime does not possess a regular past (future) null infinity for $\epsilon=+1$ ($\epsilon=-1$). Therefore, the radiating solution ($\epsilon=+1$) is only asymptotically flat to the future, whereas the accreting solution ($\epsilon=-1$) is only asymptotically flat to the past.

The remainder of this section is dedicated to a detailed analysis of the properties and causal structure of these spacetimes.

\subsection{Mass and energy flux through null infinity}

For spherically symmetric spacetimes one defines the Misner-Sharp mass\footnote{This is a well-defined quasi-local mass in general relativity. It is reasonable to expect that it remains an appropriate quasi-local mass in Einstein-Maxwell-dilaton theory, since it just adds two matter fields directly coupled to each other, but this has not been established. We thank V. Faraoni for this cautionary remark.} as~\cite{Hayward:1994bu}
\be
{\cal M}_{MS}(u,r) = \frac{R}{2}\left(1-g^{\mu\nu} \partial_\mu R\, \partial_\nu R\right)\,.
\ee
Recall that $R$ is the areal radius $R^2=r(r-2D(u))$.
By taking the large radius limit one obtains the Bondi-Sachs mass
\be
{\cal M}_{BS}(u) = \lim_{R\to\infty}{\cal M}_{MS}(u,r) = M(u)-\epsilon D(u) D'(u)\,.
\ee
This expression can also be read off from the subleading term in the expansion of $g_{uu}$ in inverse powers of $R$ [see Eq.~\eqref{A_function_areal}]. This quantity is occasionally referred to as the Vaidya mass in the literature.

The Bondi-Sachs mass receives a contribution from $D(u)$, in addition to the expected contribution from $M(u)$, and---as we show below---for non-static vacuum solutions it diverges as $\epsilon u\to-\infty$.

As expected, the rate at which the mass changes, i.e., the energy loss or gain (depending on the choice of $\epsilon$), is equal to (minus) the total flux of energy through null infinity,
\be
\frac{d{\cal M}_{BS}}{du} = -\epsilon \left[D'(u)\right]^2 = - \lim_{R\to\infty} 4\pi R^2 (-{T^r}_u)\,.
\ee
This energy flux is entirely due to the dilaton since the contribution from the Maxwell field  to the energy-momentum tensor, ${{T^r}_u}^{(EM)}$, vanishes identically.

Finally, observe that the partial derivative in Eq.~\eqref{eq:du_huu} has the required fall-off behavior for asymptotic flatness, $\partial_u h_{uu}=O(R^{-2})$, if and only if ${\cal M}_{BS}'(u)=0$.

\subsection{Non-staticity}

The fact that the Bondi-Sachs mass is not constant is an obvious indication that these spacetimes are not static.
We now prove that these solutions do not possess a timelike Killing vector field when $D(u)$ is not constant. 

In general, a vector field $V^\mu$ is a Killing vector field if and only if it preserves the metric, $\mathsterling_V g =0$. The angular vector fields associated with spherical symmetry are
\bea
_{\left(1\right)}V^\mu &=& \{ 0, 0, 0 ,1 \} \,, \label{Killing_v_1} \\
_{\left(2\right)}V^\mu &=& \{ 0, 0, \sin\varphi  ,\cot\theta \cos\varphi\}\,, \\
_{\left(3\right)}V^\mu &=& \{ 0, 0,\cos\varphi  , - \cot\theta \sin\varphi \}  \label{Killing_v_3}\,.
\eea
Additionally, we wish to consider a timelike Killing vector field of the form $K^\mu = \{ f_1, f_2, f_3 ,f_4 \}$, where the functions $f_i$ only depend on $r$ and $u$ due to the rotational symmetry already being accounted for by Eqs.~\eqref{Killing_v_1}-\eqref{Killing_v_3}. The Killing equation $\nabla_{(\mu} K_{\nu)}=0$, with the metric in the gauge of Eq.~\eqref{eq:metricsol}, yields
\bea
f_1 \partial_r A + 2 \epsilon \partial_u f_1 &=&  f_2 \left[ \partial_u A+ \epsilon A \partial_r A \right] \,, \label{f1_final} \\
\partial_r f_1 + \partial_u f_2 &=& \epsilon f_2 \partial_r A \,, \label{eq:f2_from_f1} \\
\partial_u f_a &=& -\frac{2 D'}{r - 2 D}f_a\,, \qquad\qquad\quad\,  a=3,4\,,\\
\partial_r f_2 &=& 0\,,  \label{eq:f2=f2_u}\\
\partial_r f_a  &=&  2 \frac{r- D}{r \left(r- 2D\right)} f_a\,, \qquad\qquad  a=3,4\,, \label{eq:f3_def} \\
\cot\theta f_a &=& 0\,, \qquad\qquad\qquad\qquad\quad\;\;\,  a=3,4\,, \label{eq:fa} \\
f_1 &=& \epsilon\, A f_2 + \frac{r D'}{r - D}f_2\,. \label{eq:f1_f2}
\eea
where $A(u,r) = 1 - 2M(u)/r - 2\epsilon D'(u)$.

Equations~\eqref{eq:fa} immediately give $f_3=f_4 = 0$, and from Eq.~\eqref{eq:f2=f2_u} we obtain $f_2(u,r) = f_2 (u)$.
If we then take the derivative of Eq.~\eqref{eq:f1_f2} with respect to $r$ and equate it to Eq.~\eqref{eq:f2_from_f1} we obtain the following differential equation for $f_2(u)$,
\be
\partial_u f_2 = \frac{D D'}{\left(r- D\right)^2}f_2 \,,
\ee
which nonetheless still has to satisfy Eq.~\eqref{eq:f2=f2_u}. Consequently, we obtain $f_2 = 0$. Inserting this result back into Eq.~\eqref{eq:f1_f2} fixes $f_1=0$ as the only solution.
Thus, we have shown that these spacetimes do not possess a timelike Killing vector field, if $D(u)$ is not constant.

\subsection{The vacuum constraint\label{sec:vacuum}}

To obtain dilaton-electrovacuum solutions of the field equations~\eqref{eq:fieldeqs} one needs to satisfy the vacuum constraint~\eqref{eq:ODEforD}, which we now examine carefully.

The differential equation~\eqref{eq:ODEforD} is invariant under translations $u \to u- u_0$, and also under the combined transformations $u\to-u$ and $D\to-D$. Hence, any solution that crosses $D(u)=0$ must be an odd function around some $u=u_0$ (at which moment they actually yield a diverging mass).

In fact, a first integral of the constraint~\eqref{eq:ODEforD} can be easily found to be
\be
D'(u) - \frac{\epsilon Q^2}{4 D(u)^2} = -\epsilon c\,,
\label{first_integral}
\ee
where $c$ is a real constant. We include the factor $-\epsilon$ on the right hand side for convenience. If $c=0$, the solutions are simple power laws that we present in Appendix~\ref{App2}. However, those solutions are pathological and so we will be interested in $c\neq0$.

It is convenient to rescale $D\to Q \hat{D}$ and $u\to Q \hat{u}$, so that we can equivalently set $Q=1$ and reintroduce it later if needed,
\be
\hat{D}'(\hat{u}) = \epsilon\left(\frac{1}{4\hat{D}(\hat{u})^2} - c\right)\,.
\label{ODErescaled}
\ee
A general solution in closed form cannot be obtained explicitly. The trivial exception is the constant case, $\hat{D}(\hat{u})=1/(2\sqrt{c})$, which is a valid solution only for positive $c$. Otherwise, it is possible to express $\hat{D}(\hat{u})$ implicitly as the solution of the following equation~\cite{Gueven:1996zm}
\be
\hat{D}(\hat{u})-\frac{1}{4\sqrt{c}} \log\left( \frac{2\sqrt{c} \hat{D}(\hat{u})+1}{2\sqrt{c} \hat{D}(\hat{u})-1} \right) = -\epsilon c(\hat{u}-\hat{u}_*)\,.
\label{sol_implicit}
\ee

\begin{figure}[t]
\centering
\includegraphics[width=\textwidth]{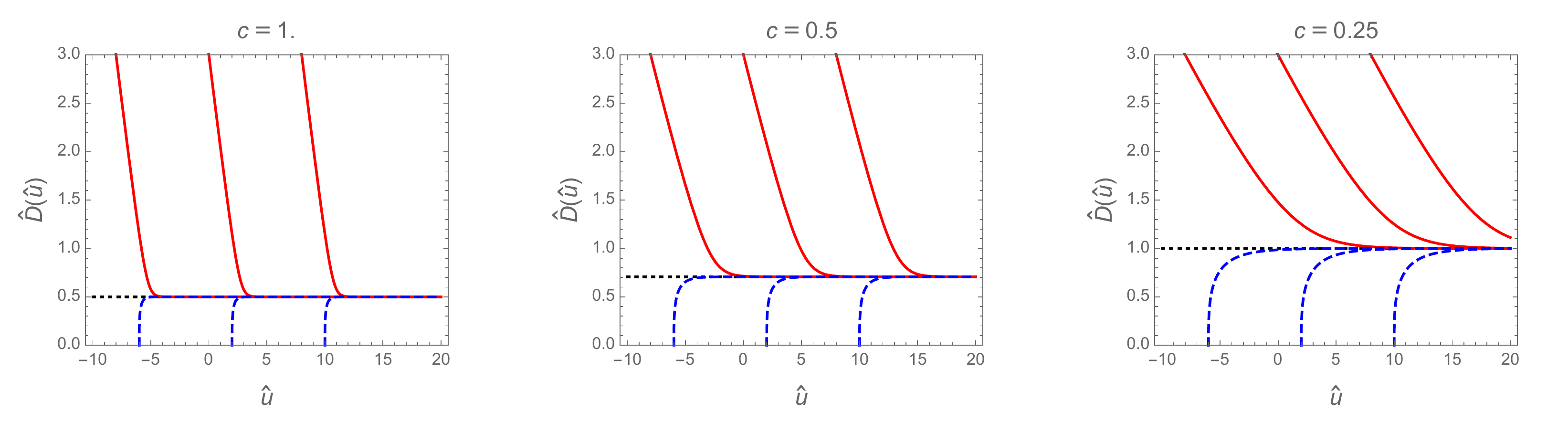}
\caption{The three classes of solutions for the function $\hat{D}(\hat{u})$, with positive $c$, and for $\epsilon=+1$. Therefore, $u$ is a retarded null coordinate. Different panels display the solutions for three choices of $c$, namely $c=1, 1/2$, and $1/4$. Class {\bf A} solutions are represented with solid (red) curves, class {\bf B} in dashed (blue) lines, and the class {\bf C} solution is the horizontal dotted (black) line. Solutions within each family differ only by a shift in $\hat{u}$.}
\label{fig:radiating}
\end{figure}

\begin{figure}[t]
\centering
\includegraphics[width=\textwidth]{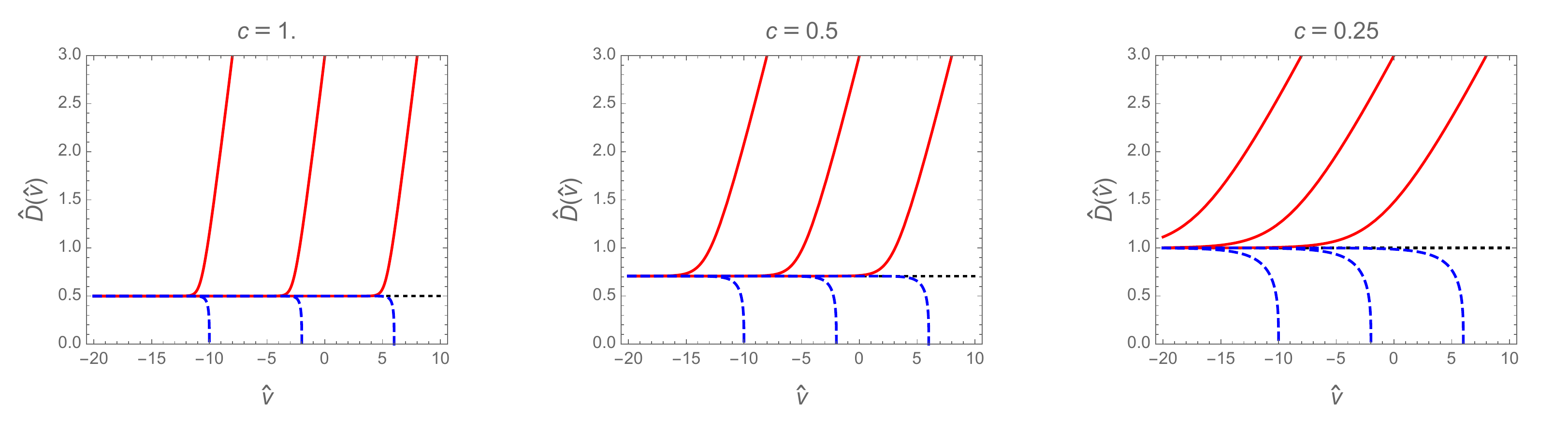}
\caption{The three classes of solutions for the function $\hat{D}(\hat{v})$, with positive $c$, and for $\epsilon=-1$. To make it clear that the null coordinate in this case is an advanced time we use $\hat{v}$ instead of $\hat{u}$. Different panels display the solutions for three choices of $c$, namely $c=1, 1/2$, and $1/4$. Same color coding and line style as in figure~\ref{fig:radiating}.}
\label{fig:accreting}
\end{figure}

The equation above is not very enlightening. Nevertheless, Eq.~\eqref{ODErescaled} defines an autonomous system and the space of solutions is straightforward to classify based on the associated vector field. For $c>0$ there are three qualitatively different classes of solutions:
\begin{enumerate}[label={\bf \Alph*}.]
\item Those for which $\hat{D}(\hat{u})$ (exponentially) asymptotes $1/(2\sqrt{c})$ from above as $\epsilon u\to\infty$, and grows linearly as $\epsilon u\to -\infty$;
\item Those for which $\hat{D}(\hat{u})$ (exponentially) asymptotes $1/(2\sqrt{c})$ from below as $\epsilon u\to\infty$, and that hit $\hat{D}(\hat{u})=0$ at finite $\hat{u}$;
\item The constant solution, $\hat{D}(\hat{u})=1/(2\sqrt{c})$, which lies at the boundary between class {\bf A} and {\bf B} solutions.
\end{enumerate}
Distinct solutions within family {\bf A} or {\bf B} are trivially related by the aforementioned shift in the retarded/advanced coordinate.
These solutions are displayed in Figs.~\ref{fig:radiating} and~\ref{fig:accreting}. To avoid confusion, we reserve the use of the letter $u$ for the null coordinate of the radiating solutions ($\epsilon=+1$), and instead employ the coordinate $v$ for accreting solutions ($\epsilon=-1$).

As $c\to0^{+}$ the class {\bf A} solutions are sent off to infinity and for $c<0$ only type {\bf B} solutions survive. These solutions reach $D(u)=0$ at finite $u$ and therefore present the same kind of pathology as the power law solutions for $c=0$ (discussed in Appendix~\ref{App2}). Therefore, we shall discard the case $c<0$ as well\footnote{One might consider truncating at some finite retarded/advanced time, $\epsilon u>\epsilon u_{cut}$, where the signature of the spacetime is asymptotically preserved, and then patching it with other spacetimes possessing physically reasonable behavior. If this is achieved, the problems associated with the null singularity at $u=0$ are avoided altogether.}. This leaves the $c>0$ solutions---and among them only those within class {\bf A}---as the physically relevant spacetimes.

Inspection of equation~\eqref{first_integral}---or alternatively a quick look at Figs.~\ref{fig:radiating} and~\ref{fig:accreting}---shows that class {\bf A} solutions interpolate between a nearly constant solution when $-\epsilon u\to\infty$ and, in the opposite asymptotic end, a linearly decreasing (growing) behaviour as $D(u)\to\infty$ in the radiating (accreting) case.

The parameter $c$ controls not only the electric-to-dilaton charge ratio of the static solution, but also the relaxation timescale for approaching the static limit~\cite{Gueven:1996zm,Lu:2014eta}. The constant solution has $D(u)=D_0=Q/(2\sqrt{c})$ and a small perturbation, $D(u)\simeq D_0(1+\varepsilon(u))$ with $\varepsilon\ll D_0$, takes the form
\be
\varepsilon(u) \propto e^{-\epsilon u/u_s}\,, \qquad {\rm where} \quad u_s= \frac{Q}{4c^{3/2}}\,.
\ee
Therefore, the static limit is approached in an exponential fashion, and the timescale $u_s$ is also determined by $c$. Alternatively, this can also be inferred from Eq.~\eqref{sol_implicit}.

In contrast, the approach to the quasi-linear regime $D(u)\sim -\epsilon c$, which emerges as $D\to\infty$, is inverse polynomial. By writing $D(u)\simeq -\epsilon c u + \delta(u)$ with $|\delta|\ll |cu|\to\infty$, one finds
\be
\delta(u)= -\epsilon \frac{Q^2}{4c^2 u}\,.
\ee

In conclusion, the physically relevant class {\bf A} solutions are parametrised by two constants, the electric charge $Q$ and the number $c$ which is related to the asymptotic mass through $M(\epsilon u\to\infty)=Q^2/(2D_0)=Q\sqrt{c}$. A trivial shift in the dilaton field $\phi\to\phi_0$ and a translation of the retarded/advanced null coordinate $u\to u-u_0$ produces 2 more parameters.

\subsection{Singularity structure}

An analysis of the singularity structure of the solutions~\eqref{GYsolution} reveals that they are everywhere regular except at $r=2D(u)$, $r=0$ and $D(u)=0$. This is indicated by the expressions for the Ricci scalar, 
\be
{\cal R}=\frac{2r (1+2 c) D(u)^3 - 2 \left(2 c r^2+Q^2\right)D(u)^2 - Q^2 r D(u) + Q^2 r^2}{r^3 D(u) (r-2 D(u))^2}\,,
\ee
and for the Kretschmann scalar,
\bea
K&\!\!\!\!=&\!\!\!\!R^{\mu\nu\rho\sigma}R_{\mu\nu\rho\sigma}=
\frac{1}{r^6 D(u)^2 (r-2 D(u))^4}
\left[ 12 r^2(1+2 c)^2 D(u)^6 -8 (2 c+1) r D(u)^5 \left(6 c r^2+5 Q^2\right) \right. \nn\\
&&+4D(u)^4 \left(12 c^2 r^4+(34 c+7) Q^2 r^2+27 Q^4\right)-4 Q^2 r D(u)^3 \left((16 c+1) r^2+49 Q^2\right) \nn\\
&&\left.+D(u)^2 \left(8 c Q^2 r^4+143 Q^4 r^2\right)-50 Q^4 r^3 D(u)+7 Q^4 r^4\right]\,.
\eea

At $r=2D(u)$ the Ricci scalar blows up like $(r-2D(u))^{-2}$ and the Kretchmann scalar diverges like $(r-2D(u))^{-4}$, unless $c=1/2$, in which case they behave as $\sim(r-2D(u))^{-1}$ and $\sim(r-2D(u))^{-3}$, respectively.

The physically relevant solutions of class {\bf A}, which necessarily have $c>0$, never approach $D(u)=0$; this function is larger than $Q/(2\sqrt{c})$ for all $u$. Hence, the possible singularity at $D(u)=0$ is never explored. Likewise, the singular spherical surface $r=0$ is not part of the spacetime; it is cut off by the curvature singularity at $r=2D(u)$, which is therefore the only physical curvature singularity of the spacetime.

Let us now study the character of this singular surface, defined by  $S(u,r)\equiv r-2D(u)=0$.
It is timelike, spacelike or null depending on whether its normal vector $\partial_\mu S$ has positive, negative or zero norm, respectively. The norm of the vector perpendicular to $S$ is computed by
\be
\left. g^{\mu\nu}\partial_\mu S\, \partial_\nu S\right|_{S=0} = 1-\frac{Q^2}{2D(u)^2}+2\epsilon D'(u) = 1-2c\,.
\ee
The norm is constant for any solution of~\eqref{eq:ODEforD}. We conclude that the singularity cannot change character as time evolves. Moreover, the singularity is timelike if $c<1/2$, null if $c=1/2$, and spacelike if $c>1/2$.

\subsection{Apparent and event horizons}

The concept of an event horizon in a dynamical setting is troublesome due to its teleological nature~\cite{Faraoni:2013aba}. Much more useful is the notion of apparent horizon, a hypersurface where the expansion of radial outgoing null geodesics vanishes (while the expansion of radial ingoing null geodesics is negative)\footnote{This definition refers to future apparent horizons. For past apparent horizons one must interchange ``outgoing'' with ``ingoing'', and ``negative'' with ``positive''.}. This is given by $\nabla_\mu R \nabla^\mu R = g^{\mu\nu}\partial_\mu(r^2B)\partial_\nu(r^2B)=0$. Using coordinates such that the line element reads as~\eqref{eq:metricsol}, the apparent horizon is determined by the following condition,
\be
A(2B+r\partial_r B)-2\epsilon r\partial_u B=0\,.
\ee
This reduces to $A=0$ when $B(u,r)=1$ but in general the surface where $A(u,r)$ vanishes will not correspond to the apparent horizon.
Note that for $\epsilon=+1$ the condition yields the past apparent horizon, while setting $\epsilon=-1$ gives the future apparent horizon.

In terms of the functions $D(u)$ and $M(u)$ the apparent horizon is indicated by the solution of
\be
(r-2M)\left(1-\frac{D}{r}\right) + 2\epsilon DD'=0\,.
\ee
It is straightforward to show that the apparent horizon is spacelike when $c>1/2$, and that it becomes degenerate with the (null) curvature singularity $r = 2D(u)$ when $c = 1/2$. For $c<1/2$ there is no apparent horizon.

In what concerns event horizons, the situation differs significantly from the static GMGHS solutions.
Consider outgoing radial light rays to the past and future (for $\epsilon=+1$ and $\epsilon=-1$, respectively). These satisfy
\be
\frac{dr}{du} = -\epsilon \left(\frac{1}{2}-\frac{M}{r}\right) + D'\,.
\ee
For early/late times (corresponding to $-\epsilon u\to\infty$) the scalar charge function is in a quasi-linear regime, $D(u)\sim -\epsilon c u$. Since $M$ is inversely proportional to $D$, these light rays behave asymptotically like
\be
r(u)\sim -\epsilon u \frac{1+2c}{2}\,.
\ee
This should be compared with the approximately linear growth of the singularity $r_s(u)=2D(u)\sim -2\epsilon c u$. The two slopes exactly match for the critical case $c=1/2$, and in the supercritical case, $c<1/2$, null geodesics can stream out to $r\to\infty$ unimpeded.

However, when $c>1/2$ the radial location of the singular surface grows superluminally. Therefore, in this case all outgoing light rays (as well as ingoing null geodesics) terminate at the spacelike singularity. We conclude that the spacetime does not have a light-like past (future) null infinity for $\epsilon=+1$ ($\epsilon=-1$) and hence a past (future) event horizon is not even defined.

In contrast, for late/early times (corresponding to $\epsilon u\to\infty$) the function $D(u)$ asymptotes a constant, so in this limit the spacetime behaves just like the static GMGHS black holes. In particular, it will feature a future (past) event horizon at $r=2M(u\to\epsilon\infty)=constant$.

\subsection{Causal structure}

Under a coordinate transformation $(u,r)\to(u,v)$ such that
\be
dv=du+2\epsilon \frac{W(u,r)}{A(u,r)} dr\,,
\label{eq:Kruskalcoord}
\ee
the line element can be brought to the Kruskal form
\be
ds^2= -A(u,r(u,v))du\,dv+r(u,v)^2 d\Omega^2\,,
\ee
which is appropriate to obtain Carter-Penrose diagrams upon conformal compactification. Unfortunately, the condition \eqref{eq:Kruskalcoord} does not have an integral in closed form, so an explicit expression for $r=r(u,v)$ is out of reach.

Nevertheless, we can extract the causal behavior from all the results above, without recourse to Kruskal coordinates.

From the above discussion it is clear that the causal structures of these time-dependent dilaton-electrovacuum spacetimes for $c=1/2$ and $c<1/2$ are the same as that of the extremal and overextremal GMGHS static spacetimes, respectively. Therefore the Carter-Penrose diagrams are {\em exactly} the same as those of the GMGHS static spacetimes with $D=M$ and $D>M$, respectively Figs.~\ref{fig:subfig_extremal} and~\ref{fig:subfig2}.

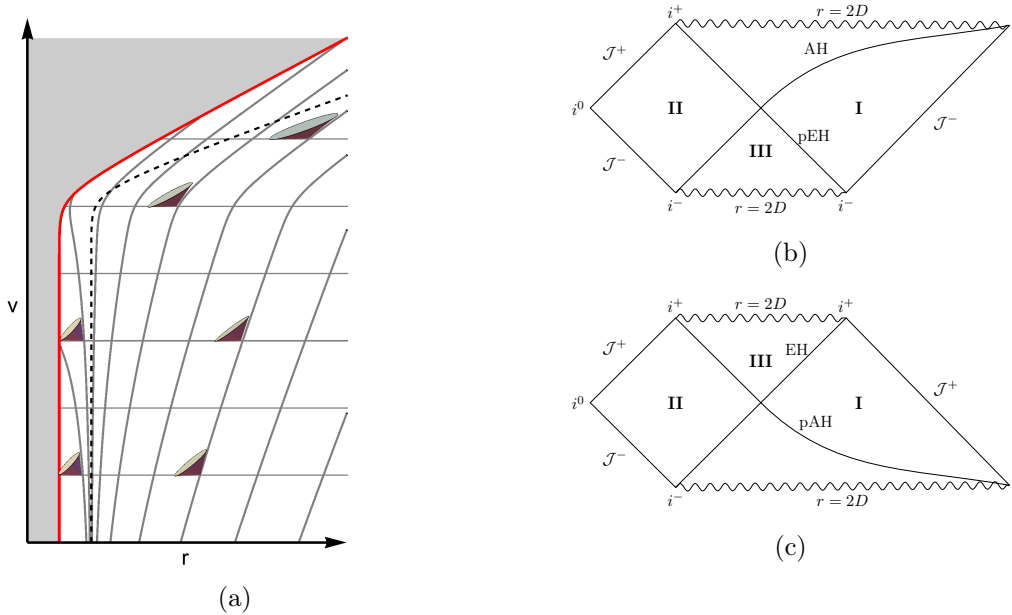
\begin{figure}[t]
\centering
	\begin{subfigure}[t]{0.5\textwidth}
		\qquad
		\includegraphics[width=4.5cm]{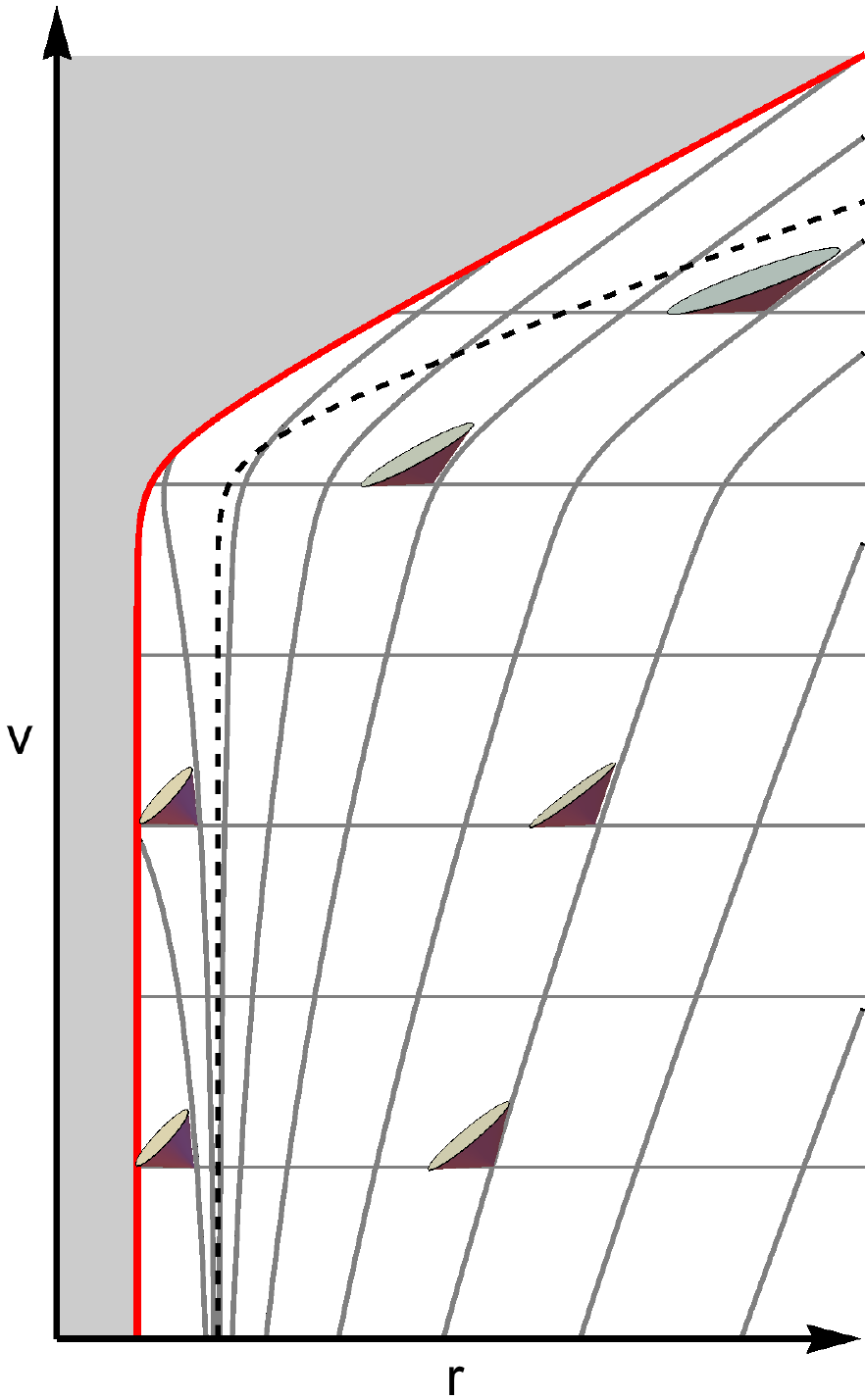}
		\caption{}
		\label{fig:STdiag}
	\end{subfigure}
	\quad
	\begin{subfigure}[b]{0.4\textwidth}
		\centering
		\resizebox{\linewidth}{!}{
			\begin{tikzpicture}
			\node (I)    at ( 2,0)  {} ;
			\node (II)   at (-2,0)  {};
			\node (III)  at (0, 1) {};
			\node (IV)   at (0,-1) {};

			\path  
			(II) +(90:2)  coordinate[label=90:$i^+$]  (IItop)
			+(-90:2) coordinate[label=-90:$i^-$] (IIbot)
			+(0:2)   coordinate                  (IIright)
			+(180:2) coordinate[label=180:$i^0$] (IIleft)
			;
			\draw (IIleft) -- 
			node[midway, above left]    {$\cal{J}^+$}
			(IItop) --
			(IIright) -- 
			(IIbot) --
			node[midway, below left]    {$\cal{J}^-$}   
			(IIleft) -- cycle;
			
			\path 
			(I) +(27:4.3)  coordinate[label=90:$ $]  (Itop)
			+(-90:2) coordinate[label=-90:$i^-$]  (Ibot)
			+(180:2) coordinate (Ileft)
			;
			\draw  
			(Itop) -- 
			node[midway, below right]    {$\cal{J}^-$}
			(Ibot) -- 
			node[midway, above] {\quad\;pEH}
			(Ileft) 
			;
			
			\path 
			(II) +(-65:2.2)  coordinate[label=90:]  (IIbot2);
			
			
			\draw[decorate,decoration=snake] (IItop) -- (Itop)
			node[midway, above, inner sep=2mm] {$r=2D$};
			
			\draw[decorate,decoration=snake] (IIbot) -- (Ibot)
			node[midway, below, inner sep=2mm] {$r=2D$};
			
			\draw[line width=0.2mm]
		(IIright) to[out=45, in=190, looseness=1] node [above, midway, pos=.25, inner sep=3mm] {AH}  (Itop);

			\node at(2.3,0)  {\large{\bf I}};
			\node at(-2,0)  {\large{\bf II}};					
			\node at(0,-1)  {\large{\bf III}};
			
			\end{tikzpicture}
		}
		\caption{}
		\label{fig:subfigB}
		
		\vspace{0.2cm}
		
		\resizebox{\linewidth}{!}{
			\begin{tikzpicture}
			\node (I)    at ( 2,0)  {} ;
			\node (II)   at (-2,0)  {};
			\node (III)  at (0, 1) {};
			\node (IV)   at (0,-1) {};

			\path  
			(II) +(90:2)  coordinate[label=90:$i^+$]  (IItop)
			+(-90:2) coordinate[label=-90:$i^-$] (IIbot)
			+(0:2)   coordinate                  (IIright)
			+(180:2) coordinate[label=180:$i^0$] (IIleft)
			;
			\draw (IIleft) -- 
			node[midway, above left]    {$\cal{J}^+$}
			(IItop) --
			(IIright) -- 
			(IIbot) --
			node[midway, below left]    {$\cal{J}^-$}   
			(IIleft) -- cycle;
			
			\path 
			(I) +(-27:4.3)  coordinate[label=90:$ $]  (Ibot)
			+(90:2) coordinate[label=90:$i^+$]  (Itop)
			+(180:2) coordinate (Ileft)
			;
			\draw  
			(Ibot) -- 
			node[midway, above right]    {$\cal{J}^+$}
			(Itop) -- 
			node[midway, above] {EH\;\;\;}
			(Ileft) 
			;
		
			\path 
			(II) +(-65:2.2)  coordinate[label=90:]  (IIbot2);
			
			
			\draw[decorate,decoration=snake] (IIbot) -- (Ibot)
			node[midway, below, inner sep=2mm] {$r=2D$};
			
			\draw[decorate,decoration=snake] (IItop) -- (Itop)
			node[midway, above, inner sep=2mm] {$r=2D$};
			
			\draw[line width=0.2mm]
			(IIright) to[out=-45, in=170, looseness=1] node [above, midway, pos=.25, inner sep=3mm] {pAH}  (Ibot);
			
			\node at(2.3,0)  {\large{\bf I}};
			\node at(-2,0)  {\large{\bf II}};					
			\node at(0,1)  {\large{\bf III}};
		
			\end{tikzpicture}
		}
		\caption{}
		\label{fig:subfigC}		
	\end{subfigure}
	\caption{Causal structure of the G\"uven-Y\"or\"uk solutions for the undercritical case, $c>1/2$. Panel (a) shows an accreting ($\epsilon=-1$) spacetime diagram adopting ingoing Eddington-Finkelstein coordinates $\{v,r\}$. The thick solid (red) line represents the singularity location and the dashed (black) line indicates the future apparent horizon. Horizontal thin (grey) lines are ingoing radial null geodesics, while the remaining curved thin (grey) lines are outgoing radial light rays. The region to the left of the singularity curve is not part of the spacetime. Panel (b) displays the corresponding Carter-Penrose diagram. Panel (c) displays the time-reversal of (b). The event horizon and the apparent horizon are indicated by EH and AH, respectively, while pEH and pAH denote past event and apparent horizons.}
	\label{fig:GYundercrit}
\end{figure}

Let us then focus on the underextremal case, $c>1/2$. For concreteness, consider the solution using an ingoing null coordinate, i.e., $\epsilon=-1$. This case is displayed explicitly in figure~\ref{fig:GYundercrit}.
As we discussed above, both the singularity and the future apparent horizon are spacelike. Moreover, all future-directed light rays necessarily fall into the singularity, signalling the spacetime does not have a regular future null infinity. In the far past ($v\to-\infty$) the line element smoothly connects with a static GMGHS black hole solution, thus yielding the Carter-Penrose diagram shown in figure~\ref{fig:subfigB}.

For $\epsilon=+1$, the results are similar in all respects, except that ``past'' and ``future'' are interchanged, i.e., it is time-reversed. Thus one obtains the Carter-Penrose diagram of figure\ref{fig:subfigC}.
From the point of view of an observer living in region {\bf I}, these solutions naturally describe a time-dependent solution that evolves from a Big Bang-like singularity to an asymptotically flat stationary black hole.
In the quasi-linear regime $D(u)\simeq-\epsilon c u\to\infty$, the line element~\eqref{eq:metricsol} becomes
\be
ds^2 \simeq -(1+2c) du^2 -2\epsilon du dr + r(r+2\epsilon c u)d\Omega^2\,.
\ee
By reverting back to Schwarzschild coordinates $u=t-\epsilon r/(1+2c)$ and rescaling $t\to\tau/\sqrt{1+2c}$, $r\to \sqrt{1+2c}\rho$, this metric takes a simple form,
\be
ds^2 \simeq -d\tau^2 +d\rho^2 + \rho^2\left(1+2\epsilon c \frac{\tau}{\rho}\right)d\Omega^2\,,
\ee
which shows that, in the quasi-linear regime, the spacetime is expanding (or contracting, depending on the choice of $\epsilon=\pm1$) anisotropically.

\section{Time-dependent solutions sourced by null dust}\label{sec:NullDust}

In the spirit of Vaidya's approach~\cite{Vaidya:1951zza,Vaidya:1951zz}, we now consider the effects of introducing generic time-dependence in the physical parameters characterising the static solutions (mass and charges). In particular, it is of interest to determine what kind of matter sources such solutions and whether it conforms to standard energy conditions.
To this end we add an additional term to Lagrangian~\eqref{lagrangian},
\be
{\cal L}_{\rm total} = {\cal L} + {\cal L}_{\rm mat}\,,\label{lagrangian_mat}
\ee
which gives rise to an extra contribution to the energy-momentum tensor,
\be
T_{\mu\nu}^{\rm (total)} \equiv  T_{\mu\nu} + T_{\mu\nu}^{\rm (mat)}\,, \qquad\quad {\rm with} \quad
T_{\mu\nu}^{\rm (mat)} = -\frac{2}{\sqrt{-g}} \frac{\partial {\cal L}_{\rm mat}}{\partial g^{\mu\nu}}\,.
\label{FluidTmunu}
\ee

The simplest way to bring about such a source in the Einstein equation is to relax condition~\eqref{eq:ODEforD}. This just turns on a nonvanishing $\{uu\}$ component in the tensor $T^{\rm (mat)}_{\mu\nu}$, thus introducing a null dust component, in addition to the gauge field and the dilaton. This null dust satisfies the standard energy conditions (weak, strong, dominant) as long as the left hand side of~\eqref{eq:ODEforD} is non-negative~\cite{Kuchar:1990vy,Hawking:1973uf}, i.e.,
\be
\mu \equiv 2 D(u)^3 D''(u)+\epsilon Q^2 D'(u)\geq 0\,.
\label{eq:EC}
\ee
It can be verified through an analysis similar to the one carried out in Ref.~\cite{Aniceto:2015klq} that this condition on the energy-momentum tensor of the the null dust component is sufficient to guarantee the energy conditions on the total stress-energy tensor are automatically obeyed.

\begin{figure}[!t]
	\centering
	\begin{subfigure}[b]{0.35\textwidth}
		\centering
		\resizebox{\linewidth}{!}{
		\begin{tikzpicture}
		\node(I) at (0,0) {};
		\node(II) at (-2,0) {};
		
		\path 
		(I) + (0:2) coordinate coordinate[label=0:$i^0$] (Iright)
		(I) +(180:2)  coordinate (Ileft)
		(I) + (-90:2)  coordinate[label=0:$i^-$] (Idown)
		(II) + (-90:2) coordinate (IIdown)
		(II) + (90:4) coordinate[label=0:$\;i^+$] (IItop);
		
		\filldraw[draw=black,bottom color=gray, top color=white] (1.3,0.7) -- (-1.4,-2) -- (-2,-2) -- (-2,-1.4) -- (0.7,1.3);
		
		\draw (Idown) --
		node[midway, below right]    {$\cal{J}^-$}
		node[midway, above, sloped] {$r=\infty$}
		(Iright);
		\draw (IItop) -- 
		node[midway, below, sloped] {$r=\infty$}
		node[midway, above right]    {$\cal{J}^+$}
		(Iright);
		\draw (Idown) --
		(Ileft);
		
		\path
		(II) + (90:0.7) coordinate (IImed);
		
		\draw[decorate,decoration=snake] (IIdown) -- (IItop)
		node[midway, above, sloped, outer sep=2mm] {$r=2D$};
		
		\draw[decorate,decoration=snake] (IIdown) -- (Idown)
		node[midway, below, sloped, outer sep=1mm] {$r=2D$};

		\draw (Idown) to[out=135, in=45, looseness=1]
		node[midway, above left]    {AH} 
		(IIdown);

		\end{tikzpicture}
		}
		\caption{\centering White hole radiating and turning into naked singularity.}
		\label{fig_dyn:WH_sing}
	\end{subfigure}
	\qquad
	\begin{subfigure}[b]{0.5\textwidth}
		\centering
		\resizebox{\linewidth}{!}{
			\begin{tikzpicture}
			\node (I)    at ( 2,0)  {} ;
			\node (II)   at (-2,0)  {};
			\node (III)  at (0, 1) {};
			\node (IV)   at (0,-1) {};
			\node (V)   at (2,1) {};
			
			\path  
			(II) +(90:2)  coordinate[label=90:$i^+$]  (IItop)
			+(-90:2) coordinate[label=-90:$i^-$] (IIbot)
			+(0:2)   coordinate                  (IIright)
			+(180:2) coordinate[label=180:$i^0$] (IIleft)
			;
			\draw (IIleft) -- 
			node[midway, above left]    {$\cal{J}^+$}
			(IItop) --
			(IIright) -- 
			node[midway, above, sloped] {AH}
			(IIbot) --
			node[midway, below left]    {$\cal{J}^-$}   
			(IIleft) -- cycle;
			
			\path 
			(I) +(65:2.2)  coordinate[label=90:$i^+$]  (Itop)
			+(-90:2) coordinate[label=-90:$i^-$]  (Ibot)
			+(180:2) coordinate (Ileft)
			+(12:2.4)   coordinate[label=0:$i^0$]  (Iright)
			;
			\draw  
			(Itop) -- 
			node[midway, above right]    {$\cal{J}^+$}
			(Iright) -- 
			node[midway, below right]    {$\cal{J}^-$}
			(Ibot) -- 
			(Ileft) 
			;

			\filldraw[draw=black,bottom color=white, top color=gray] (0.85,2) -- (3.45,-0.45) -- (2.65,-1.3) -- (-0.75,2); 
			
			\path 
			(II) +(-65:2.2)  coordinate[label=90:]  (IIbot2);
			
			\draw (IIbot2) -- 
			node[midway, below, sloped] {EH\;\;}
			(Itop);
			
			\draw[decorate,decoration=snake] (IItop) -- (Itop)
			node[midway, above, inner sep=2mm] {$r=2D$};
			
			\draw[decorate,decoration=snake] (IIbot) -- (Ibot)
			node[midway, below, inner sep=2mm] {$r=2D$};
			
			\path 
			(IIright) + (0.5,0.5)  coordinate[label=90:]  (IIIcent);
			
			\draw[line width=0.2mm]
			(IIright) to
			(IIIcent);
			
			\draw[line width=0.2mm]
			(IIright) + (0.5,0.5) to[out=40, in=200, looseness=1] 
			(V);
						
			\end{tikzpicture}
		}
		\caption{Black hole accreting null dust.}
		\label{fig_dyn:BH_null_dust}
	\end{subfigure}
	\\
\vspace{0.3cm}
	\begin{subfigure}[b]{0.3\textwidth}
		\centering
		\resizebox{\linewidth}{!}{
			\begin{tikzpicture}
			
			\node(I) at (0,0) {};
			\node(II) at (-2,0) {};
			
			\path 
			(I) + (0:2) coordinate[label=0:$i^0$] (Iright)
			(I) +(180:2)  coordinate (Ileft)
			(I) + (90:2)  coordinate[label=0:$\;i^+$] (Itop)
			(II) + (-90:4) coordinate[label=0:$i^-$] (IIdown)
			(II) + (90:2) coordinate (IItop);
			
			\filldraw[draw=black,bottom color=white, top color=gray] (1.3,-0.7) -- (-1.4,2) -- (-2,2) -- (-2,1.4) -- (0.7,-1.3);
			
			\draw (Itop) --
			node[midway, above right]    {$\cal{J}^+$}
			node[midway, below, sloped] {$r=\infty$}
			(Iright);
			\draw (IIdown) -- 
			node[midway, above, sloped] {$r=\infty$}
			node[midway, below right]    {$\cal{J}^-$}
			(Iright);
			\draw (Itop) --
			node[midway, below, sloped] {EH}
			(Ileft);

			
			\draw[decorate,decoration=snake] (IIdown) -- (IItop)
			node[midway, above, sloped, outer sep=2mm] {$r=2D$};
			
			\draw[decorate,decoration=snake] (IItop) -- (Itop)
			node[midway, above, sloped, outer sep=1mm] {$r=2D$};
			
			\draw (IItop) 
			to[out=-45, in=-135, looseness=1]
			node[midway, below left] {AH}
			(Itop);
			
			\end{tikzpicture}
		}
		\caption{Naked singularity turning into a black hole by accreting null dust.}
		\label{fig_dyn:naked_sing_BH}
	\end{subfigure}
	\qquad
	\begin{subfigure}[b]{0.5\textwidth}
		\centering
		\resizebox{\linewidth}{!}{
			\begin{tikzpicture}
			
			\node (I)    at ( 2,0)  {} ;
			\node (II)   at (-2,0)  {};
			\node (III)  at (0, 2) {};
			\node (IV)   at (0,-2) {};
			\node(V) at (2,2) {};
			
			\path
			(V) + (-135:1.7)  coordinate (Vmid);
			
			\path 
			(III) + (0:1.3) coordinate (IIImid);

			

			\path
			(I) + (90:4) coordinate[label=90:$i^+$]  (Vtop);
			
			\path
			(V) + (3,-1) coordinate[label=0:$i^0$] (Vright);
			
			\path  
			(II) +(90:2)  coordinate[label=90:$i^+$]  (IItop)
			+(-90:2) coordinate[label=-90:$i^-$] (IIbot)
			+(0:2)   coordinate                  (IIright)
			+(180:2) coordinate[label=180:$i^0$] (IIleft);

			\draw (IIleft) -- 
			node[midway, above left]    {$\cal{J}^+$}
			(IItop) --
			node[midway, below, sloped] { }
			(IIright) -- 
			node[midway, above, sloped] { }
			(IIbot) --
			node[midway, below left]    {$\cal{J}^-$}   
			(IIleft) -- cycle;
			
			\path 
			(I) +(90:2)  coordinate  (Itop)
			+(-90:2) coordinate[label=-90:$i^-$]  (Ibot)
			+(180:2) coordinate (Ileft)
			+(0:2)   coordinate (Iright)
			;
			
			\path
			(Itop) + (90:0.707107) coordinate (boxtopright);
			
			\path
			(boxtopright) + (-45:3.32843) coordinate (boxbotright);
			
			\path
			(boxbotright) + (-135:1) coordinate (boxbotleft);
			
			\path
			(boxbotleft) + (135:3.32843) coordinate (boxtopleft);

			\filldraw[draw=black,bottom color=white, top color=gray] (Itop) -- (boxtopright) -- (boxbotright) -- (boxbotleft) -- (boxtopleft);
			
			\draw  (Ileft) -- 
			node[midway, below, sloped] { }
			(Itop);
			\draw (Ibot) -- 
			node[midway, above, sloped] {pEH}
			(Ileft) ;
			
			\draw (Ibot) --
			node[midway, below right]    {$\cal{J}^-$}
			(Vright) --
			node[midway, above right]    {$\cal{J}^+$}
			(Vtop);
			
			\draw[decorate,decoration=snake] (IItop) -- (Itop)
			node[midway, above, inner sep=2mm] {$r=2D$};
			
			\draw[decorate,decoration=snake] (IIbot) -- (Ibot)
			node[midway, below, inner sep=2mm] {$r=2D$};
			
			\draw[decorate,decoration=snake] (Itop) -- (Vtop)
			node[midway, above, sloped, inner sep=2mm] {$r=2D$};
			
			\end{tikzpicture}
		}
		\caption{Black hole accreting null dust with negative energy density and turning into a naked singularity.}
		\label{fig_dyn:BH_null_dust_violation}
	\end{subfigure}	
	\caption[Penrose diagrams of ...]{Some examples of Carter-Penrose diagrams corresponding to dynamical spacetimes sourced by null dust described in the main text. The event horizon and the apparent horizon are indicated by EH and AH, respectively, while pEH denotes a past event horizon.} 
	\label{fig:Penrose3}
\end{figure}
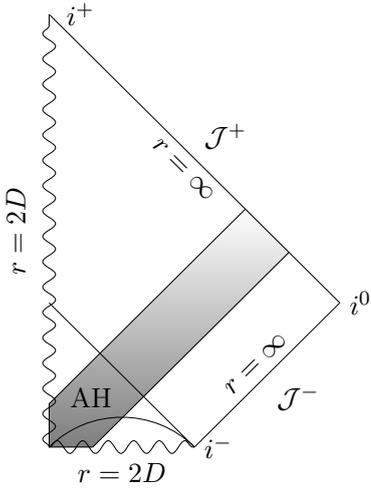
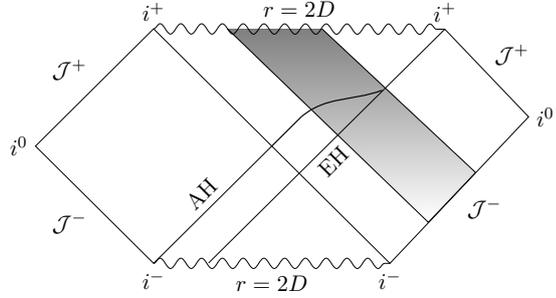
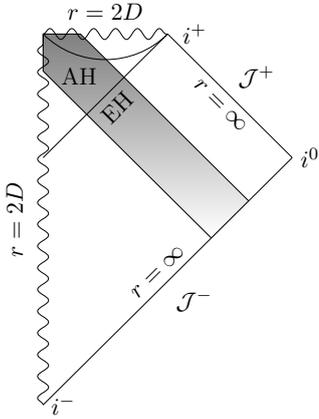
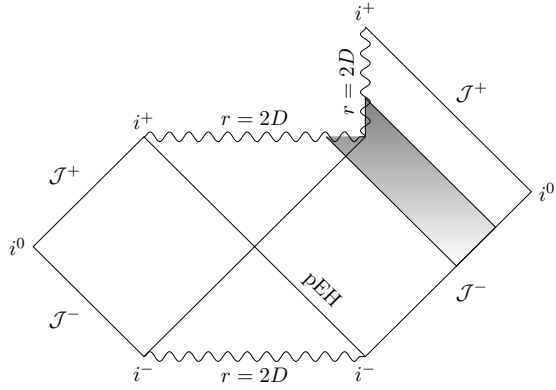

One can easily devise such time-dependent non-vacuum solutions interpolating between asymptotically static solutions.
A concrete example is given by the following function:
\be
D(u)= \frac{Q}{\sqrt{2}}\left(1+\frac{\epsilon}{8} \arctan (u/Q)\right)\,. \label{dilaton_charge_null_dust}
\ee 
For the radiating case ($\epsilon=+1$) this interpolates between a nearly static solution with $M>D$ at early times (featuring an apparent horizon) and a nearly static solution with $M<D$ at late times (a naked singularity). This should be reversed for the accreting case.
However, this {\em does not} correspond to a black hole radiating and leaving behind a naked singularity. To see why, recall that for the radiating case $u$ is an outgoing coordinate so it only penetrates the {\em past} apparent horizon. Hence, the initial state describes a white hole instead of a black hole (see figure~\ref{fig_dyn:WH_sing}).
It would be desirable to employ instead the accreting solution, where $u$ is an ingoing coordinate that penetrates the future apparent horizon, but then to have $M$ decreasing we would need to violate energy conditions.

More generally, to construct an accreting spacetime ($\epsilon=-1$) evolving from a nearly static black hole in the far past into a nearly static naked singularity in the far future we need the constant $c$ to be promoted to a function $c(u)$ that interpolates between $c(u\to-\infty)>1/2$ and $c(u\to\infty)<1/2$. For this to occur, the function $c(u)$ must be decreasing, at least in some interval. In terms of the function $D(u)$ this translates into
\be
0>c'(u)=\frac{d}{du}\left[D'(u)+\frac{Q^2}{4D(u)^2}\right]=\frac{1}{2D(u)^3}\left[ 2D(u)^3D''(u)-Q^2D'(u) \right]\,.
\ee
From~\eqref{eq:EC}, such an evolution of a black hole into a naked singularity implies that the null dust must have negative energy density.

To conclude, we can construct exact solutions describing a white hole evolving into a naked singularity or a naked singularity evolving into a black hole, by radiating or accreting null dust with positive energy density. To evolve a black hole into a naked singularity requires the null dust to violate standard energy conditions. Thus, we find no indication for violation of weak cosmic censorship in low-energy effective string theory.

\section{Matching dynamical solutions along null hypersurfaces}\label{sec:matching}

The fact that the dynamical vacuum solutions we have been studying have explicit dependence on the retarded or advanced null coordinate $u$ enables us to easily match two such spacetimes across a lightlike thin shell, using the standard Barrab\'es-Israel junction conditions~\cite{Barrabes:1991ng,Poisson:2004}. This is an interesting setup to test weak cosmic censorship in an analytic framework. It can be viewed as a limit of compressing the null dust component of the previous section to an infinitely thin pulse. 

We relegate the details of the spacetime-matching formalism to Appendix~\ref{App3} and present here only the main outputs of those calculations. Equality of the induced metrics on the null hypersurface imposes
\be
\left[ D(u) \right] = 0\,,
\label{junction_condition_D}
\ee
where the notation $[{\cal A}]\equiv {\cal A}_+ - {\cal A}_-$ denotes the jump of a given quantity ${\cal A}$ across the selected hypersurface.
The junction conditions allow for a discontinuity of the so-called transverse curvature, and, if nonvanishing, this indicates the presence of a surface stress-energy tensor. This turns out to be of the null dust form,
\be
S^{\alpha \beta} \equiv \mu k^\alpha k^\beta\,,
\ee
with
\be
\mu = \frac{r D(u) \left[D'(u) \right] - \epsilon  \left(r-D(u)\right) \left[M(u)\right] }{4 \pi r^2 (r-2D(u))} \,,\label{eq:matching_energy_density}
\ee
and where $k^\alpha$ is a null vector tangent to the generators of the hypersurface.

In the following we will apply these junction conditions to the cases in which the interior is either a time-dependent GY solution or a static GMGHS black hole spacetime, and the exterior corresponds to a naked singularity. These are the cases of interest to test weak cosmic censorship.

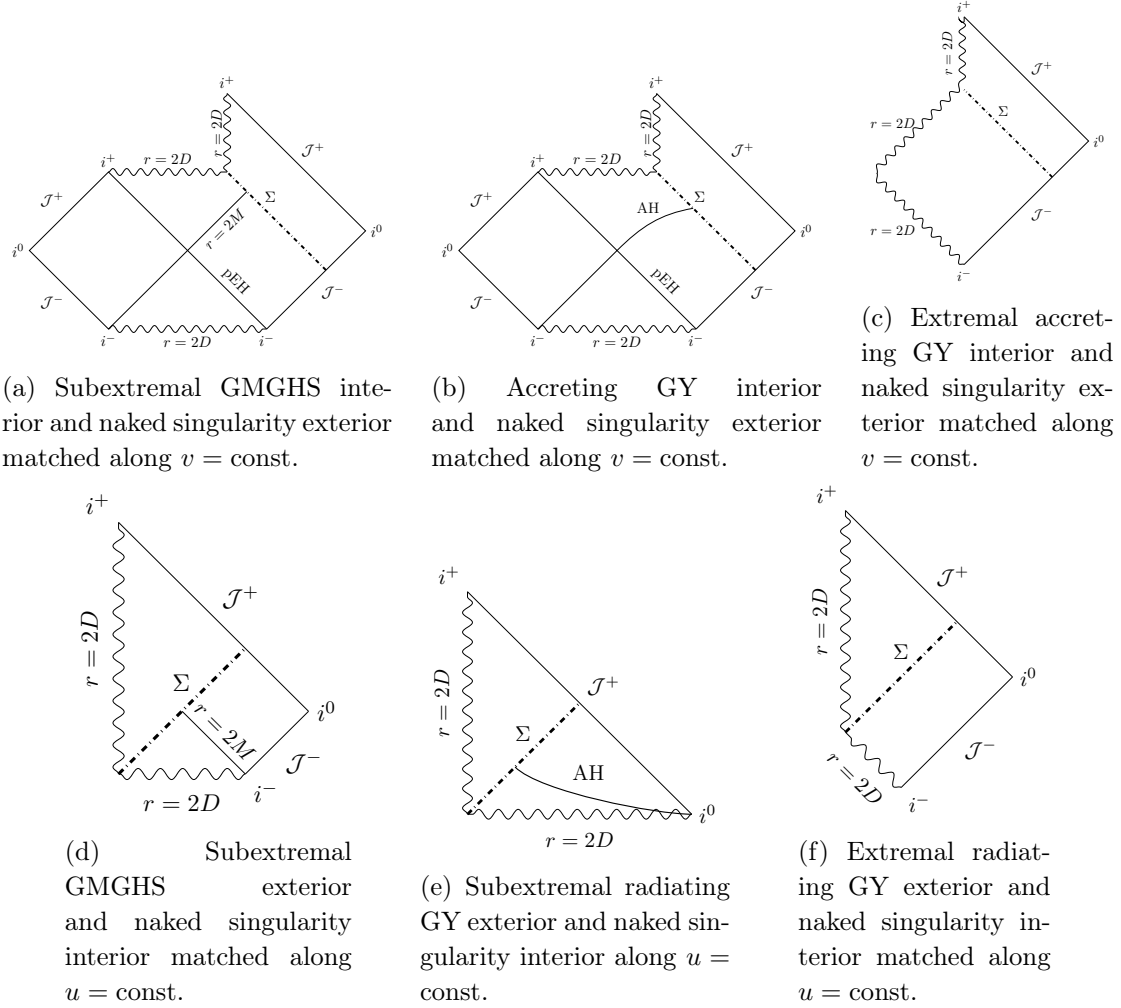
\begin{figure}[!t]
	\centering	
	\begin{subfigure}[b]{0.34\textwidth}
		\centering
		\resizebox{\linewidth}{!}{
			\begin{tikzpicture}
			
			\node (I)    at ( 1,0)  {} ;
			\node (II)   at (-2,0)  {};
			\node (III)  at (0, 2) {};
			\node (IV)   at (0,-2) {};
			\node(V) at (2,2) {};
			
			\path
			(I) + (90:4) coordinate[label=90:$i^+$]  (Vtop);
			
			\path
			(Vtop) + (-45:4.94975) coordinate[label=0:$i^0$] (Vright);
			
			\path  
			(II) +(90:2)  coordinate[label=90:$i^+$]  (IItop)
			+(-90:2) coordinate[label=-90:$i^-$] (IIbot)
			+(0:2)   coordinate                  (IIright)
			+(180:2) coordinate[label=180:$i^0$] (IIleft)
			;
			
			\path 
			(IIbot) + (0:4)  coordinate[label=-90:$i^-$]  (Ibot);

			\draw (IIleft) -- 
			node[midway, above left]    {$\cal{J}^+$}
			(IItop) --
			node[midway, below, sloped] {}
			(IIright) -- 
			node[midway, above, sloped] {}
			(IIbot) --
			node[midway, below left]    {$\cal{J}^-$}   
			(IIleft) -- cycle;
			
			\path 
			(I) +(90:2)  coordinate  (Itop);
			
			\path
			(Itop) +(-45:3.5355) coordinate (Lmidright);
			
			\path
			(IIright) + (45:2.12132) coordinate (midhyp);
			
			\draw (Ibot) -- 
			node[midway, above, sloped] { pEH }
			(IIright) ;
			
			\draw (Ibot) --
			node[midway, below right]    {$\cal{J}^-$}
			(Vright) --
			node[midway, above right]    {$\cal{J}^+$}
			(Vtop);
			
			\draw (IIright)  -- 
			node[midway, below, sloped] { $r=2M$ }
			(midhyp);
			
			\draw[decorate,decoration=snake] (IItop) -- (Itop)
			node[midway, above, inner sep=2mm] {$r=2D$};
			
			\draw[decorate,decoration=snake] (IIbot) -- (Ibot)
			node[midway, below, inner sep=2mm] {$r=2D$};
			
			\draw[decorate,decoration=snake] (Itop) -- (Vtop)
			node[midway, above, sloped] {$r=2D$};
			
			\draw[dashdotted, line width=0.5mm]
			(Lmidright) to node [right, pos=.75, inner sep=3mm] {$\Sigma$}  (Itop);
			
			\end{tikzpicture}
		}
		\caption{Subextremal GMGHS interior and naked singularity exterior matched along $v= \mathrm{const.}$}
		\label{fig_matching:subextremal_static_advanced}  
	\end{subfigure}
	\quad
	\begin{subfigure}[b]{0.34\textwidth}
		\centering
		\resizebox{\linewidth}{!}{
			\begin{tikzpicture}
			
			\node (I)    at ( 1,0)  {} ;
			\node (II)   at (-2,0)  {};
			\node (III)  at (0, 2) {};
			\node (IV)   at (0,-2) {};
			\node(V) at (2,2) {};
			
			\path
			(I) + (90:4) coordinate[label=90:$i^+$]  (Vtop);
			
			\path
			(Vtop) + (-45:4.94975) coordinate[label=0:$i^0$] (Vright);
			
			\path  
			(II) +(90:2)  coordinate[label=90:$i^+$]  (IItop)
			+(-90:2) coordinate[label=-90:$i^-$] (IIbot)
			+(0:2)   coordinate                  (IIright)
			+(180:2) coordinate[label=180:$i^0$] (IIleft)
			;
			
			\path 
			(IIbot) + (0:4)  coordinate[label=-90:$i^-$]  (Ibot);

			\draw (IIleft) -- 
			node[midway, above left]    {$\cal{J}^+$}
			(IItop) --
			node[midway, below, sloped] { }
			(IIright) -- 
			node[midway, above, sloped] { }
			(IIbot) --
			node[midway, below left]    {$\cal{J}^-$}   
			(IIleft) -- cycle;
			
			\path 
			(I) +(90:2)  coordinate  (Itop);
			
			\path
			(Itop) +(-45:3.5355) coordinate (Lmidright);
			
			\path
			(IIright) + (45:2.12132) coordinate (midhyp);

			\draw (Ibot) -- 
			node[midway, above, sloped] {pEH}
			(IIright) ;
			
			\draw (Ibot) --
			node[midway, below right]    {$\cal{J}^-$}
			(Vright) --
			node[midway, above right]    {$\cal{J}^+$}
			(Vtop);
			
			\path
			(midhyp) + (-45:0.6) coordinate (hypbot);
			
			\draw[line width=0.2mm]
			(IIright) to[out=45, in=190, looseness=1] 
			node [above, midway, pos=.45, inner sep=3mm] {AH}  
			(hypbot);

			\draw[decorate,decoration=snake] (IItop) -- (Itop)
			node[midway, above, inner sep=2mm] {$r=2D$};
			
			\draw[decorate,decoration=snake] (IIbot) -- (Ibot)
			node[midway, below, inner sep=2mm] {$r=2D$};
			
			\draw[decorate,decoration=snake] (Itop) -- (Vtop)
			node[midway, above, sloped] {$r=2D$};
			
			\draw[dashdotted, line width=0.5mm]
			(Lmidright) to node [right, pos=.75, inner sep=3mm] {$\Sigma$}  (Itop);
			
			\end{tikzpicture}
		}
		\caption{Accreting GY interior and naked singularity exterior matched along $v= {\rm const.}$}
		\label{fig_matching:subextremal_dyn_advanced}
	\end{subfigure}		
	\quad
	\begin{subfigure}[b]{0.22\textwidth}
		\centering
		\resizebox{\linewidth}{!}{
			\begin{tikzpicture}
			
			\node (I)    at (1,0)   {};
			\node (II)    at (1,0)   {};
			
			\path
			(II) +(90:2.4)  coordinate  (IItop)
			+(-90:2.4) coordinate[label=-90:$i^-$] (IIbot)
			+(0:2.4) coordinate (IIright)
			+(180:2.4) coordinate[label=180:$ $] (IIleft)
			;			
			
			\path
			(IIright) + (45:1.4) coordinate[label=0:$i^0$] (Vright);

			\path
			(IItop) + (90:2) coordinate[label=90:$i^+$] (Vtop);
			
			\draw[decorate,decoration=snake] (IIbot) -- 
			node[midway, below left] {$r=2D$}
			(IIleft) --
			node[midway, above left] {$r=2D$}
			(IItop) -- 
			node[midway, above, sloped, outer sep=2mm] {$r=2D$}			
			(Vtop);
			
			\draw (Vtop) -- 
			node[midway, above right]    {$\cal{J}^+$}
			(Vright) --
			node[midway, below right] {$\cal{J}^-$}
			(IIbot) ;		
			
			\draw[dashdotted, line width=0.5mm]
			(IIright) to node [right, pos=.75, inner sep=3mm] {$\Sigma$}  (IItop);

			\end{tikzpicture}
		}
		\caption{Extremal accreting GY interior and naked singularity exterior matched along $v=\mathrm{const.}$}
		\label{fig_matching:extremal_advanced}  
	\end{subfigure}
	\\
	\begin{subfigure}[b]{0.25\textwidth}
		\centering
		\resizebox{\linewidth}{!}{
			\begin{tikzpicture}

			\node (I)    at (-1,0)   {};
			
			\path
			(I) + (-90:1) coordinate (II);
			
			\path
			(II) + (45:2.8) coordinate (IIright)
			+ (90:4) coordinate[label=120:$i^+$]  (IItop)
			+ (0:2) coordinate[label=-45:$i^-$] (IIbot);

			\draw[decorate,decoration=snake] (IIbot) --
			node[midway, below, sloped, outer sep=2mm] {$r=2D$}
			(II);
			
			\draw[decorate,decoration=snake] (II)--
			node[midway, above, sloped, outer sep=2mm] {$r=2D$}
			(IItop);
		
			\path
			(IIbot) + (45:1.41) coordinate[label=0:$i^0$] (IIbotright);
			
			\draw
			(IIbot) --
			node[midway, below right]    {$\cal{J}^-$}
			(IIbotright);

			\draw
			(IIbotright) -- 
			node[midway, above right]    {$\cal{J}^+$}
			(IItop);			

			\draw[dashdotted, line width=0.5mm]
			(II) to node [midway, above, outer sep=2mm] {$\Sigma$}  (IIright);
			
			\path
			(IIbot) + (135:1.41) coordinate (IIcenter);
			
			\draw (IIbot) -- 
			node[midway, above, sloped] {  \small{$r=2M$}}
			(IIcenter);

			\end{tikzpicture}
		}
		\caption{Subextremal GMGHS exterior and naked singularity interior matched along $u= \mathrm{const.}$}
		\label{fig_matching:subextremal_retarded} 
	\end{subfigure}
	\qquad
	\begin{subfigure}[b]{0.27\textwidth}
		\centering
		\resizebox{\linewidth}{!}{
			\begin{tikzpicture}

			\node (I)    at (-1,0)   {};
			
			\path
			(I) + (-90:1) coordinate (II);
			
			\path
			(II) + (45:2.8) coordinate (IIright)
			+ (90:4) coordinate[label=120:$i^+$]  (IItop)
			+ (0:2) coordinate (IIbot);

			\draw[decorate,decoration=snake] (II)--
			node[midway, above, sloped, outer sep=2mm] {$r=2D$}
			(IItop);
			
			\path
			(IIbot) + (0:2) coordinate[label=0:$i^0$] (IIbotright);
			
			\draw[decorate,decoration=snake] (IIbotright) --
			node[midway, below, sloped, outer sep=2mm] {$r=2D$}
			(II);

			\draw
			(IIbotright) -- 
			node[midway, above right]    {$\cal{J}^+$}
			(IItop);			
			
			\draw[dashdotted, line width=0.5mm]
			(II) to node [midway, above, outer sep=2mm] {$\Sigma$}  (IIright);
			
			\path
			(IIbot) + (135:1.41) coordinate (IIcenter);
			
			\path
			(II) + (45:1.2) coordinate (IIAH);

			\draw[line width=0.2mm]
			(IIAH) to[out=-45, in=180, looseness=0.5] 
			node [above, midway, pos=.45, inner sep=3mm] {AH}  
			(IIbotright);								
			
			\end{tikzpicture}
		}
		\caption{Subextremal radiating GY exterior and naked singularity interior along $u= \mathrm{const.}$}
		\label{fig_matching:GY_retarded} 
	\end{subfigure}	
	\qquad
	\begin{subfigure}[b]{0.22\textwidth}
		\centering
		\resizebox{\linewidth}{!}{
			\begin{tikzpicture}
			
			\node (I)    at (-1,0)   {};
			
			\path
			(I) + (-90:1) coordinate (II);
			
			\path
			(II) + (45:2.8) coordinate (IIright)
			+ (90:4) coordinate[label=120:$i^+$]  (IItop)
			+ (-45:1.41) coordinate[label=-45:$i^-$] (IIbot);

			\draw[decorate,decoration=snake] (IIbot) --
			node[midway, below, sloped, outer sep=2mm] {$r=2D$}
			(II);
			
			\draw[decorate,decoration=snake] (II)--
			node[midway, above, sloped, outer sep=2mm] {$r=2D$}
			(IItop);
			
			\path
			(IIbot) + (45:2.82) coordinate[label=0:$i^0$] (IIbotright);
			
			\draw
			(IIbotright) -- 
			node[midway, above right]    {$\cal{J}^+$}
			(IItop);			
			
			\draw
			(IIbot) --
			node[midway, below right]    {$\cal{J}^-$}
			(IIbotright);
			
			\draw[dashdotted, line width=0.5mm]
			(II) to node [midway, above, outer sep=2mm] {$\Sigma$}  (IIright);
						
			\end{tikzpicture}
		}
		\caption{Extremal radiating GY exterior and naked singularity interior matched along $u=\mathrm{const.}$}
		\label{fig_matching:extremal_retarded} 
	\end{subfigure}	
	
	\caption[Penrose diagrams]{Carter-Penrose diagrams for global solutions obtained by matching along a null hypersurface. All of these matchings yield a spacetime with the same global asymptotic structure of Minkowski, except for figure~\ref{fig_matching:GY_retarded}.} 
	\label{fig:Penrose2}
\end{figure}

The different matchings that will be considered are summarised in figure~\ref{fig:Penrose2}. For all of them we wish to determine whether the standard energy conditions are satisfied, which in the present case involving only null dust, correspond simply to $\mu \geq 0$.
To proceed, it is useful to consider separately matchings performed along the hypersurface $v=\mathrm{const.}$ (i.e., along the advanced null coordinate, which corresponds to $\epsilon = -1$) and matchings along the hypersurface $u= \mathrm{const.}$ (i.e., along the retarded null coordinate, which corresponds to $\epsilon=+1$).

%

\subsection{Matching along the advanced null coordinate}

For the matching using the advanced null coordinate $v$ we consider $\epsilon=-1$ and the setups shown in Figs.~\ref{fig_matching:subextremal_static_advanced},~\ref{fig_matching:subextremal_dyn_advanced} and~\ref{fig_matching:extremal_advanced}. We start by analysing the subextremal and extremal GY accreting solution as the interior, i.e., we consider that at the hypersurface $v=\mathrm{const.}$ we have $c_- \geq 1/2$.
Additionally, for the exterior solution we are interested in an overextremal GY accreting solution, i.e., $c_+ < 1/2$ which implies $M_+(v) < D(v)$.\footnote{Note that for class {\bf A} solutions---to which we restrict---one necessarily has $D>Q/(2\sqrt{c})$ and this imposes $M=Q^2/(2D)<2cD$. So if $c\leq1/2$ one retrieves $M<D$, but this is not guaranteed if $c>1/2$.}
Thus, plugging Eqs.~\eqref{first_integral} and~\eqref{junction_condition_D} in Eq.~\eqref{eq:matching_energy_density}, and furthermore assuming $[Q]=0$ in order for the null matter shell to be electrically uncharged (as in Section~\ref{sec:NullDust}), yields
\be
4\pi R(v,r)^2 \mu(v) = D(v) \left[c\right] + \frac{r-2D(v)}{2r} \left[M(v)\right] = D(v) \left[c\right] < 0\,.
\label{eq:adv_dyn}
\ee
Therefore, regardless of the value of $v$ at which the matching is performed, we have that $\mu(v) < 0$, i.e., it does not satisfy the energy conditions. If we do not insist on $[Q]=0$, but instead assume the interior has $M_-(v)>D(v)$ (which for a dynamical spacetime is distinct from being subextremal), we still obtain $\mu(v) < 0$ since then $[M(v)]<0$.

The situations in which the interior and/or the exterior are static spacetimes arise as particular cases of Eq.~\eqref{eq:adv_dyn}, when the mass parameters and the dilaton charge are time-independent and $c_\pm = M_\pm /(2D)$.

To conclude, we have shown that it is not possible, through the accretion of a lightlike thin shell of uncharged dust, to turn any of the GY dynamical black holes into a naked singularity unless the matter content of the thin shell does not obey standard energy conditions. Consequently, weak cosmic censorship is not violated in this setting.

\subsection{Matching along the retarded null coordinate}

We shall now study matchings along a hypersurface $u=\mathrm{const.}$ For this, we use the radiating solutions ($\epsilon=+1$) and consider the setups presented in Figs.~\ref{fig_matching:subextremal_retarded},~\ref{fig_matching:GY_retarded} and~\ref{fig_matching:extremal_retarded}. As before, we have that $\left[D(u)\right] =0$, $D(u) > M_+(u)$ and $c_- \geq 1/2 >c_+$. We consider a subextremal exterior (in this case this carries the subscript $-$) and a dynamical naked singularity interior (denoted with the subscript $+$). By plugging Eq.~\eqref{first_integral} into Eq.~\eqref{eq:matching_energy_density}, and assuming again $[Q]=0$ to recover electrically neutral null shells, we obtain
\be
4\pi R(u,r)^2 \mu(u)  = - \left(  D(u) \left[c\right] + \frac{r-2D(u)}{2r} \left[M(u)\right] \right)
= - D(u) \left[c\right] \,, \label{eq:rtd_dyn}
\ee
which is positive. This result is expected since Eqs.~\eqref{first_integral} and~\eqref{eq:matching_energy_density} imply $\mu$ is proportional to $\epsilon$. Hence, the expression for $\mu(u)$ flips the overall sign with respect to the expression for $\mu(v)$ obtained in the previous section.

We conclude that for these matchings along a hypersurface $u=\mathrm{const.}$, the standard energy conditions on the matter content of the thin shell are always satisfied. It is important to stress, however, that these matchings do not represent a violation of weak cosmic censorship since the configurations preceding the outburst of radiation are not regular to begin with. Referring to Figs.~\ref{fig_matching:subextremal_retarded},~\ref{fig_matching:GY_retarded} and~\ref{fig_matching:extremal_retarded} these would correspond respectively to a white hole, a Big-Bang-like singularity and a naked null singularity.

\section{Non-stationary dyonic solutions of the Einstein-Maxwell-axion-dilaton system}\label{sec:axion}

The Einstein-Maxwell-dilaton system expressed in the Lagrangian~\eqref{lagrangian}, is a particular case of a more general class of theories in which the scalar and the gauge field are coupled through the term $e^{-2a\phi}F^2$, i.e., with strength determined by the dilaton coupling constant $a$. 
So far we have restricted our attention to $a=1$ and for this special value the theory arises as a truncation of 4D low-energy effective heterotic string theory. The bosonic part of the full low-energy effective Lagrangian is 
\be
{\cal L} =\frac{\sqrt{-g}}{16\pi}\left[R - 2(\nabla\phi)^2 - \frac{1}{2}e^{4\phi}(\nabla\chi)^2 - e^{-2\phi}F^2 - \chi F\, {}^\star F\right] \,,\label{lagrangian2}
\ee
where $\chi$ is a pseudoscalar---the axion---and ${}^\star F^{\mu\nu}\equiv \frac{1}{2\sqrt{-g}}\epsilon^{\mu\nu\rho\sigma}F_{\rho\sigma}$ denotes the Hodge dual of the field strength.
One recovers the theory described by~\eqref{lagrangian} when the axion field is consistently set to zero.

The Einstein-Maxwell-axion-dilaton system exhibits an $SL(2,\mathbb{R})$ symmetry at the classical level, and this can be used as a solution generating technique. This symmetry group contains the standard discrete electric-magnetic duality, which gets enhanced to the full $SL(2,\mathbb{R})$ when $a=1$. This is a consequence of the Lagrangian~\eqref{lagrangian2} being equivalently expressed as~\cite{Shapere:1991ta}
\be
{\cal L} =\frac{\sqrt{-g}}{16\pi}\left[R - \frac{|\nabla\lambda|^2}{2\lambda_2^2} + \frac{i}{4}\left(\lambda F_+^2 - \overline{\lambda} F_-^2\right)\right] \,,\label{lagrangian3}
\ee
where the axion and the dilaton have been combined in a complex-valued scalar field $\lambda \equiv \lambda_1 + i\lambda_2 \equiv \chi + i e^{-2\phi}$, and $F_\pm \equiv F\pm i\, {}^\star F$ are the self- and anti self-dual field strengths.
The equation of motion for the axidilaton $\lambda$, as well as the Einstein and Maxwell equations (augmented with the Bianchi identity), are invariant under the combined transformations (see for example Ref.~\cite{Schwarz:1992tn})
\begin{flalign}
& \lambda \to \frac{a\lambda+b}{c\lambda+d}, \qquad\quad
\textrm{with} \quad
 \left(
  \begin{array}{cc}
    a & b \\ 
    c & d \\ 
  \end{array}
 \right)
\in SL(2,\mathbb{R})\,, \nn\\
& F_+^{\mu\nu} \to (c\lambda+d)F_+^{\mu\nu} \,,  \qquad  F_-^{\mu\nu} \to (c\overline{\lambda}+d)F_-^{\mu\nu} \,, \label{duality}\\
& g_{\mu\nu} \to g_{\mu\nu}\,. \nn
\end{flalign}
Hence, the $SL(2,\mathbb{R})$ transformation~\eqref{duality} maps vacuum solutions of the equations of motion into distinct vacuum solutions. Nevertheless, note that it leaves the metric untouched.

The discrete electric-magnetic duality, which maps the dilaton according to $\phi\to-\phi$, is then obtained as the particular transformation $\lambda\to-1/\lambda$ (i.e., $c=-1=-b,\; a=d=0$) when the axion vanishes. In this case, the Maxwell equation and the Bianchi identity,
\be
\nabla_\mu \left( \chi {}^\star F^{\mu\nu} + e^{-2\phi} F^{\mu\nu} \right)=0\,, \qquad\quad
\nabla_\mu {}^\star F^{\mu\nu}=0\,,
\ee
get interchanged, but more generally they transform into linear combinations of the two equations.

One can then use such $SL(2,\mathbb{R})$ transformations to generate dyonic solutions of theory~\eqref{lagrangian2}, for example by taking a purely magnetic solution as a seed (which therefore has vanishing axion field). This was accomplished for static solutions in Ref.~\cite{Shapere:1991ta}. 
The procedure does not assume stationarity of the solutions so it can be applied also to the G\"uven-Y\"or\"uk solutions. The end result is a time-dependent dyonic spacetime with the same line element,
\be
 ds^2 =-\left(1-\frac{2M(u)}{r}-2\epsilon D'(u)\right) du^2 - 2\epsilon du dr + r^2 \left(1-\frac{2D(u)}{r}\right) d\Omega^2\,,
\ee
but the electromagnetic field strength and the dilaton are transformed, while the axion is nonvanishing:
\bea
 F &=& -\frac{Q_e}{r^2} du\wedge dr + Q_m \sin\theta\, d\theta\wedge d\varphi\,,\\
 e^{-2\phi}&=&\frac{r M(u) \left(r M(u)-Q^2\right)}{\left(Q_e^2-r M(u)\right)^2+Q_m^2 Q_e^2}\,, \\
 \chi &=& -\epsilon \frac{Q_m Q_e \left(Q_m^2+Q_e^2-2r M(u)\right)}{r M(u) \left(r M(u)-2Q_e^2\right)+Q_e^2 \left(Q_m^2+Q_e^2\right)}\,.
\eea
The total charge appearing in the metric components is related to the electric charge $Q_e$ and the magnetic charge $Q_m$ through $Q^2=Q_e^2+Q_m^2$.
The same constraints derived in the EMd case are required to obtain time-dependent dyonic vacuum solutions without additional matter fields, namely
\be
2M(u)D(u) = Q^2= {\rm const.}\,, \qquad\qquad
2 D(u)^3 D''(u)+\epsilon Q^2 D'(u)=0\,. \nn
\ee

Given that the line element for these time-dependent dyonic spacetimes is identical to that of the G\"uven-Y\"or\"uk solutions, the causal structure is also the same.

\section{Discussion}\label{sec:conc}

In this paper we investigated time-dependent, spherically symmetric solutions of the Einstein-Maxwell-axion-dilaton system with specific dilaton coupling $a=1$, as occurs in four dimensional low-energy effective string theory. These spacetimes have not attracted as much attention as their Einstein-(massless)scalar counterparts, but the existence of such exact solutions, valid in a dynamical regime, is valuable. In order to better understand their nature, we conducted an extensive analysis of their causal structure, thus filling a gap in the literature.
In addition, we generalised the dilaton-electrovacuum radiating solutions previously found by G\"uven and Y\"or\"uk to include (i) null dust sources and (ii) a nontrivial axion field, which occurs whenever both electric and magnetic charges are turned on. In the first case we were able to perform various tests of the weak cosmic censorship conjecture and it was shown that, within this class of solutions, a regular black hole cannot evolve into a naked singularity if the null dust satisfies the standard energy conditions.

The causal structure obtained from the G\"uven-Y\"or\"uk solutions differs significantly from the one observed in numerical studies of the collapse of a charged scalar field in dilaton gravity~\cite{Borkowska:2011zh}. In any case, strong similarities should not be expected since the matter content is actually different and also because the simulations show an evolving electric charge, whereas the solutions we studied have constant charge.

We have focused on black hole (or nakedly singular) solutions but these geometries can be useful for studying radiating stars if one considers matching with an appropriate (non-vacuum) interior solution. The resulting spacetime would have non-vanishing charge $Q\neq0$ in order to be dynamical, with the exception of the limit in which Roberts' spacetime is recovered as the exterior solution (i.e. $Q=M=0$, and a linear dilaton charge $D(u)=\sigma u$).

The proof that our dynamical solutions sourced by physically acceptable null dust do not evolve spacetimes with a spacelike singularity into spacetimes with a timelike singularity also addresses the {\em strong} version of cosmic censorship, i.e., whether an observer falling towards the singularity can see it for some time before he hits it. This would occur only if the original spacelike singularity could become timelike, at least for a moment, but as we have shown, in this case the null dust violates standard energy conditions.

We have seen that the G\"uven-Y\"or\"uk spacetimes fall short of being truly asymptotically flat, since they miss part of the asymptotic structure of Minkowski (this refers to a regular and light-like null infinity, either past or future). This raises the question of whether a Birkhoff-like theorem can be proved for Einstein-Maxwell-dilaton theory. I.e., if a regular dilaton-electrovacuum solution is spherically symmetric and asymptotically flat, does that imply it is static?

An interesting extension of this work would be to obtain solutions with non-constant charge $Q$, sourced by null dust. This would allow to have an exact description of the formation of black holes (or naked singularities) in the context of low-energy effective string theory. A class of such solutions was already discussed in Ref.~\cite{Aniceto:2015klq}, although the matter supporting those configurations is not of the simple null dust form. In any case, a varying electric charge will require the additional matter to be charged.

Can one obtain analogous time-dependent, spherically symmetric vacuum solutions in EMd with arbitrary value for the dilaton coupling $a$? This appears to be a challenging task. The technique we used to rederive the G\"uven-Y\"or\"uk solutions does not generalise straightforwardly to $a\neq1$. Even if one allows the electric charge to depend on advanced/retarded time, $Q(u)$, spherically symmetric non-static solutions require $|a|=1$ if the electric field is assumed to be Coulombian, $F_{ur}\propto r^{-2}$, as is shown in Appendix~\ref{App1}. It seems likely that solutions with $|a|\neq1$ must have $r^2 F_{ur}$ with non-trivial dependence on both $u$ and $r$.
Nevertheless, it is striking that exact radiating solutions are known only in the case in which S-duality is present. Could this coincidence have a deeper meaning?

It is worth mentioning that time-dependent solutions that are asymptotically AdS have been found in three-dimensional gravity with a conformal scalar field~\cite{Ayon-Beato:2015ada}. On the other hand, the 4D solutions with supergravity-inspired potentials discussed in Refs.~\cite{Lu:2014eta,Zhang:2014sta,Zhang:2014dfa,Fan:2016yqv}, which also have (anti-)de Sitter asymptotics, can be obtained from their asymptotically flat counterparts by a simple procedure laid out in~\cite{Mignemi:2009ui}.
In principle, this process can also be used to induce (anti-)de Sitter asymptotics in the new time-dependent dyonic solutions obtained above, while simultaneously generating a nontrivial potential for the scalar fields.

\appendix
\section{Solutions with generic dilaton coupling constant}\label{App1}

A broader class of Einstein-Maxwell-dilaton theories includes a coupling between the dilaton and the gauge field whose strength is governed by a real parameter $a$,
\be
{\cal L} =\frac{\sqrt{-g}}{16\pi}\left[R-2(\nabla\phi)^2-e^{-2a\phi}F^2\right]\,. \label{lagrangian_a}
\ee
For $a=1$ we recover the particular theory considered in the bulk of this article.
The field equations derived from Eq.~\eqref{lagrangian_a} read 
\begin{subequations}\label{eq:fieldeqs_a}
\bea
&&\nabla_\mu \left( e^{-2a \phi} F^{\mu \nu} \right) = 0\,,\label{MaxwellEOM_a}\\
&&\nabla^2 \phi + \frac{a}{2}e^{-2a\phi} F^2 = 0\,,\label{dilatonEOM_a}\\
&&G_{\mu\nu} = 8\pi  T_{\mu\nu} \equiv 8\pi \left(T_{\mu\nu}^{\rm (dil)} + e^{-2a\phi}T_{\mu\nu}^{\rm (EM)} \right)\,,\label{EinsteinEOM_a}
\eea
\end{subequations}
and the energy-momentum tensors $T_{\mu\nu}^{\rm (dil)}$ and $T_{\mu\nu}^{\rm (EM)}$ were defined in Eqs.~\eqref{Tdil} and~\eqref{TEM}.
In general, the dilaton equation of motion is not independent. It is implied by the divergence of the Einstein equation, as long as the dilaton is not constant (see, e.g., section 12.1 of~\cite{Ortin:2004ms}).

Static solutions of the field equations~\eqref{eq:fieldeqs_a} were presented in Refs.~\cite{Gibbons:1987ps,Garfinkle:1990qj}. 
The electrically charged solution can be expressed in Eddington-Finkelstein coordinates according to
\begin{eqnarray}
 ds^2 &=& -A(r) du^2 - 2\epsilon\, dudr + r^2 B(r) d\Omega^2 \,, \label{metricGHS_a} \\
 F &=&-\frac{Q}{r^2} du\wedge dr\,,\label{FaradayGHS_a} \\
 e^{2a\phi(r)}&=& B(r)\,, \label{dilatonTdependent_a} 
\end{eqnarray}
with the two metric functions defined by
\be
A(r)=\left(1-\frac{r_+}{r}\right) B(r)^{\frac{1-a^2}{2a^2}}\,, \qquad
B(r)=\left(1-\frac{r_-}{r}\right)^{\frac{2a^2}{1+a^2}}\,, 
\ee
and subject to the constraint
\be
(1+a^2)Q^2 = r_- r_+\,.
\label{eq:QMDconstraint_a}
\ee
Here we are insisting that the dilaton vanishes asymptotically, $\phi_0=\lim_{r\to\infty}\phi=0$, but we remind the reader that a nonzero $\phi_0$ can be trivially generated by shifting the scalar field.

Now we look for time-dependent electric solutions of the general form
\begin{subequations}\label{eq:ansatz_a}
\begin{eqnarray}
 ds^2 &=& -A(u,r) du^2 - 2\epsilon\, dudr + r^2 B(u,r) d\Omega^2 \,, \label{metricGHS_a2} \\
 F &=&-\frac{X(u,r)}{r^2} du\wedge dr\,,\label{FaradayGHS_a2} \\
 e^{2a\phi}&=& e^{2a\phi(u,r)} \,. \label{dilatonTdependent_a2} 
\end{eqnarray}
\end{subequations}
The Maxwell equations can be readily integrated to yield
\be
e^{2a\phi(u,r)}=  \frac{B(u,r)X(u,r)}{Q} \,,
\label{eq:MaxwellConstraint_a}
\ee
where $Q$ is an arbitrary constant. Restricting to static configurations and assuming $B(u,r)\underset{r\to\infty}{\longrightarrow}1$ for asymptotic flatness, the parameter $Q$ is none other than the electric charge,
\be
Q = \lim_{r\to\infty} X(r) = \frac{1}{4\pi} \int_{S^2_\infty} e^{-2a\phi} \star F\,.
\ee

The $rr$ component of the Einstein equations is a constraint equation only involving radial derivatives,
\be
(\partial_r \phi)^2 
= -\frac{1}{2r^2\sqrt{B}}\partial_r\left(r^2\frac{\partial_r B}{\sqrt{B}}\right) \,.
\ee
Plugging this in Eq.~\eqref{eq:MaxwellConstraint_a} one obtains the following constraint:
\be
\frac{\partial_r X}{X} = - \frac{\partial_r B}{B} + a \sqrt{-\frac{2}{r^2\sqrt{B}}\partial_r\left(r^2\frac{\partial_r B}{\sqrt{B}}\right)}\,.
\ee

Next, from the $ur$ component of the Einstein equations we find (after using the above relation to eliminate $\partial_r X$)
\be
X = \frac{r^2}{Q} \left\{1-\partial_r\left[\frac{A}{2}\partial_r \left(r^2 B\right) - \epsilon r^2 \partial_u B \right]\right\}\,.
\label{repX}
\ee
This determines $X$ in terms of $A$ and $B$.

At this point we note that if one is given a static GMGHS solution~\eqref{metricGHS_a}, a family of time-dependent solutions to the field equations analysed up to now can be immediately obtained by promoting the parameters of the static solution to become functions of advanced/retarded time, $r_\pm\to r_\pm(u)$ [while still obeying~\eqref{eq:QMDconstraint_a}], and simultaneously shifting the metric function $A$ according to
\bea
 A(r) &\to& A^{\rm dyn}(u,r) \equiv \left.A(r)\right|_{r_\pm=r_\pm(u)} + \frac{2}{\partial_r\left[r^2 B^{\rm dyn}(u,r)\right]} \left[\epsilon r^2 \partial_u B^{\rm dyn}(u,r) + \xi(u) \right], \\
 B(r) &\to& B^{\rm dyn}(u,r) \equiv \left.B(r)\right|_{r_-=r_-(u)}\,,
\eea
so that the left hand side of~\eqref{repX} remains invariant. This means we are insisting that the function $X$ equals its static counterpart $X(u,r)=Q$, and in particular that the dynamical solution generated still has constant electric charge.

The procedure above introduces an arbitrary function of the advanced/retarded time, $\xi(u)$.
Inserting the transformed metric functions $A^{\rm dyn}(u,r)$ and $B^{\rm dyn}(u,r)$ in the $\theta\theta$ component of the Einstein equation determines the solution for $\xi$,
\be
\xi(u) = \frac{2 \epsilon\, r_-'(u) \left[\left(a^4-1\right) r+r_-(u)\right]}{\left(a^2+1\right)^2} \left(1-\frac{r_-(u)}{r}\right)^{\frac{a^2-1}{a^2+1}}\,.
\label{xi}
\ee
As mentioned previously, for a non-constant dilaton the scalar field equation~\eqref{dilatonEOM_a} is not independent of the Einstein equation and we find it yields the same result for $\xi$.

Now, there are only two possibilities allowing expression~\eqref{xi} to be independent of the radial coordinate, as it must be: either $a^2=1$ or $r_-'(u)=0$. In the latter case one simply recovers the static solution. For the former we obtain
\be
\xi(u) = \frac{1}{2}\epsilon\, r_-(u)r_-'(u)\,.
\ee

The only remaining nontrivial equation to check is the $uu$ component of the Einstein equation. Plugging in the expressions obtained so far for $A(u,r)$, $B(u,r)$ and $\xi(u)$ we get a nonlinear constraint required to obey the vacuum equations,
\be
r_-(u)^3 r_-''(u) + 4\epsilon Q^2 r_-'(u) = 0\,.
\ee

This procedure succeeds in generating time-dependent solutions with constant electric charge from static solutions, but only for dilaton coupling $a^2=1$. In this case the mass and dilaton charge are related to the parameters $r_\pm$ through~\cite{Garfinkle:1990qj}
\be
r_+=2M\,, \qquad 
r_-=2D\,,
\ee
so that the vacuum constraint can be equivalently expressed as
\be
2D(u)^3 D''(u) + \epsilon Q^2 D'(u) = 0\,.
\ee

\smallskip
In the above derivation we explicitly assumed that the electric charge was constant, namely $X(u,r)=Q$.
Indeed, any spherically symmetric solution of the field equations~\eqref{eq:fieldeqs_a} of the form~\eqref{eq:ansatz_a} with $X=\mathrm{const.}$ is either static or has dilaton coupling $a=\pm1$.
Nevertheless, the same conclusion holds even for the less restrictive case $X=X(u)$, for which the electric charge is {\em not} constant. The reason is that the forms of the line element~\eqref{metricGHS_a2} and of the field strength~\eqref{FaradayGHS_a2} remain invariant under the following coordinate transformation:
\be
u=\int \frac{d\tilde{u}}{\eta(\tilde{u})}\,, \qquad\qquad
r=\eta(\tilde{u}) \tilde{r}\,.
\ee
The net effect is only to change the metric functions according to
\be
A \to \frac{A}{\eta}+2\epsilon \frac{\eta'}{\eta} \tilde{r}\,, \qquad\quad
B \to \eta^2 B\,,
\ee
while the electromagnetic field strength gets rescaled,
\be
X \to \frac{X}{\eta^2}\,.
\ee
Therefore, by picking $\eta = d\tilde{u}/du = \sqrt{X(u)/Q}$ one reverts back to the case $X=\mathrm{const.}=Q$ already considered above.

\section{Power law solutions of the vacuum constraint}\label{App2}

A simple power-law solution of the vacuum constraint~\eqref{first_integral}, which we reproduce here for convenience
\be
D'(u) - \frac{\epsilon Q^2}{4 D(u)^2} = -\epsilon c\,, \nn
\ee
can be easily found when $c=0$:
\be
D(u)= \alpha u^{1/3}\,, \qquad
M(u)= \frac{2\epsilon \alpha^2}{3 u^{1/3}}\,,  \qquad {\rm with} \quad
Q^2=\frac{4\epsilon\alpha^3}{3}\,.
\label{eq:power-law}
\ee
A simple translation $u\to u-u_0$ allows us to obtain the most general solution for $c=0$, so without loss of generality we set $u_0=0$.

The last identity in Eq.~\eqref{eq:power-law} clearly implies that $\epsilon \alpha \geq 0$. This solution can only have physical significance for $\epsilon u\geq0$, otherwise the mass is negative. For the explicit power-law evolution~\eqref{eq:power-law} we have ${\cal M}(u)=\frac{\epsilon \alpha^2}{3 u^{1/3}}$ so the Bondi mass actually diverges when $u\to0$.


Besides yielding a diverging Bondi mass, this spacetime is pathological at more severe level. For sufficiently small retarded/advanced coordinate $u\ll1$, the signature of the spacetime is not Lorentzian, not even as $r\to\infty$. This is an unwanted feature for a physical isolated system and we therefore discard these particular solutions.

\section{Junction conditions for null hypersurfaces in EMd theory}\label{App3}

We begin by fixing some notation. The null hypersurface separating spacetime in an exterior region $\mathcal{V}^+$ and an interior region $\mathcal{V}^-$, which identifies the thin matter layer, will be denoted by $\Sigma$. In general, each region is parametrised by its own coordinates $x^\alpha_\pm$ for which the metric is $g^\pm_{\alpha \beta}$. However, for the simple cases we will consider the same coordinates can be used on both sides of the shell. Therefore, we will drop the indices $\pm$ in the coordinates, although the metric functions can differ in the exterior and interior regions, a priori. The two line elements then take the general form
\be
 ds_\pm^2 = g^{\pm}_{\alpha\beta}dx^\alpha dx^\beta = -A_\pm(u,r) du^2 -2 \epsilon du dr + R_\pm(u,r)^2 d\Omega^2 \label{metric_matching}\,.
\ee
The conversion to the notation used in Appendix~\ref{App1} is accomplished by making $R(u,r)^2=r^2 B(u,r)$.
Moreover, we parametrise the hypersurface through the intrinsic coordinate system $\{y^a\}$. (Henceforth, we shall use greek indices when considering the $4$-dimensional spacetime, whereas latin indices refer to the $3$-dimensional hypersurface.)
Finally, we shall use $[{\cal A}] \equiv {\cal A}^+|_\Sigma - {\cal A}^-|_\Sigma$ to represent the jump of any quantity ${\cal A}$ across the hypersurface. 

To perform the matching we follow the procedure described in Refs.~\cite{Barrabes:1991ng,Poisson:2004}. We consider the hypersurface to be at $u = \mathrm{const}$. The induced metric on the hypersurface is then given by
\be
ds_\Sigma ^2 = R(u,r)^2 d\Omega^2\,.
\label{eq:induced_metric}
\ee
The null generators of the hypersurface $\Sigma$ are labelled by the angular coordinates $\vartheta^A=(\theta,\varphi)$, with $A=\{2,3\}$. We adopt $\lambda=\epsilon r$ as the parameter along each generator, so that $\lambda$ increases to the future. The tangent vectors to the hypersurface are $e^\alpha_a = \partial x^\alpha / \partial y^a$, where $y^a = (\lambda, \vartheta^A$). Accordingly, the tangent vector to each null generator is $k^\alpha = e^\alpha_\lambda = \epsilon \, e^\alpha_r$. These vectors satisfy $k_\alpha k^\alpha = k_\alpha e^\alpha_A = 0$. Additionally, we can identify Eq.~\eqref{eq:induced_metric} with the inner product of the spacelike vectors $e^\alpha_A$ since
\be
ds_{\Sigma}^2 = \sigma_{AB}\, d\vartheta^A d\vartheta^B\,,
\ee
with $\sigma_{AB} \equiv g_{\alpha \beta}e^\alpha_A e^\beta_B$. To complete the basis of vectors which is used to describe the matching across $\Sigma$, one introduces a null vector $N^\alpha$ such that $N_\alpha k^\alpha = -1$ and $N_\alpha e^\alpha_A=0$. Thus, we obtain
\be
N^\alpha \partial_\alpha = \partial_u - \epsilon \frac{A(u,r)}{2} \partial_r \label{N_transverse_vector} \,.
\ee
With this basis of vectors defined one can now enunciate the junction conditions imposed on the metric. The intrinsic metric of the hypersurface needs to satisfy
\be
\left[ \sigma_{AB} \right] = 0\,.
\label{1st_junction_condition}
\ee
When applied to the dynamical spacetimes we are matching, which have $R(u,r)^2=r\left(r-2D(u)\right)$, this condition translates into
\be
\left[ D(u) \right] = 0\,.
\label{junction_condition_D_app}
\ee
Additionally, the components of the singular part of the stress-energy tensor, supported on the thin shell,
\be
S^{\alpha \beta} \equiv \mu k^\alpha k^\beta + j^A \left(k^\alpha e^\beta_A  + e^\alpha_A k^\beta  \right) + p\, \sigma^{AB}e^\alpha_A e^\beta_B\,,
\ee
are given by 
\bea
\mu  &=& -\frac{1}{8\pi} \sigma^{AB}\left[ C_{AB}\right] \,,\label{mu_def} \\
j^A &=& \frac{1}{8 \pi }\sigma^{AB}\left[C_{rB}\right] \,,\label{jA_def} \\
p &=& -\frac{1}{8 \pi} \left[C_{rr}\right] \,, \label{P_def}
\eea
with $C_{ab}$ corresponding to the transverse curvature of the hypersurface, $C_{ab} \equiv (\nabla_\beta N_\alpha) e^\alpha_a e^\beta_b$. The quantity $\mu$ is regarded as a mass density, $j^A$ gives the matter flow in the direction transverse to the generators and $p$ is an isotropic pressure.\footnote{It is worth noting that there is no rest frame for a lightlike shell and therefore these quantities cannot be given the absolute meaning of a shell energy density, current and pressure. To obtain the physical quantities measured by observers one needs to define a congruence of timelike geodesics, with tangent vector field $u^\alpha$, upon which the physical quantities are proportional to the bare quantities multiplied by $\left( -k_\mu u^\mu  \right)^{-1}$. Nevertheless, for our analysis the bare quantities suffice.} 

Using Eq.~\eqref{N_transverse_vector}, the transverse curvature yields
\bea
 C_{rr} &=& C_{rA} = C_{\theta \varphi} = 0 \,, \label{eq:C_rr}  \\
 C_{\theta \theta} &=& \frac{C_{\varphi \varphi}}{\sin^2\theta} = \frac{2 \partial_u \left(R^2 \right)- \epsilon A\, \partial_r \left(R^2 \right)}{4}\label{eq:C_thetatheta} \,.
\eea
Thus, we obtain $p= j^A = 0$.
From Eqs.~\eqref{junction_condition_D_app} and~\eqref{eq:C_thetatheta} we conclude that 
\be
\mu = \frac{r D(u) \left[D'(u) \right]   - \epsilon  \left(r-D(u)\right) \left[M(u)\right]   }{4 \pi r^2 (r-2D(u))} \,. \label{eq:matching_energy_density_app}
\ee

The presence of the Maxwell field and the scalar field imposes further constraints, in addition to the junction conditions obtained above from the matching of the metrics. This is necessary to guarantee that the matter fields are properly defined as distributions on the entire spacetime. Following the procedure to determine the Darmois-Israel junction conditions detailed, for example, in Ref.~\cite{Poisson:2004} we obtain, as expected, that the dilaton and the electromagnetic field strength must be continuous across the shell (the case of timelike and spacelike hypersurfaces was worked out in Ref.~\cite{AnicetoThesis})
\bea
\left[ \phi \right] &=& 0\,, \label{jump_phi} \\
\left[F_{ab}\right]  &=& 0\,. \label{Fab_function_definition}
\eea
We note that Eq.~\eqref{jump_phi} yields exactly the same result as that of the first junction condition, Eq.~\eqref{junction_condition_D_app}.\footnote{One also obtains conditions on the discontinuity of the transverse derivatives of $\phi$ and $F_{ab}$, but these can be regarded just as definitions of the dilaton and electric charge surface densities. Since they will not be used, we refrain from presenting them.}

\acknowledgments
We thank Paolo Pani for early collaboration on this project.
We are also grateful to Gabriel L. Cardoso, Roberto Emparan, Valerio Faraoni, Paolo Pani and Helvi Witek for discussions and comments on the draft.
JVR acknowledges financial support from the European Union's Horizon 2020 research and innovation programme under the Marie Sk\l{}odowska-Curie grant agreement No REGMat-2014-656882.
JVR was partially supported by the Spanish MINECO under Project No. FPA2013-46570-C2-2-P. PA acknowledges financial support from the LisMath program through the FCT fellowship PD/BD/128415/2017.

\bibliographystyle{JHEP}
\bibliography{bib_DBHsEMd}

\end{document}